\documentclass[journal=jctcce,manuscript=article,layout=twocolumn]{achemso}
\usepackage{chemformula} % Formula subscripts using \ch{}
\usepackage[T1]{fontenc} % Use modern font encodings
\usepackage{graphics}
\usepackage{makecell}
\usepackage{xcolor}
\usepackage{amsmath}
\usepackage{enumerate}
\usepackage[normalem]{ulem}
\usepackage{color, colortbl}
\newcommand{\onlinecite}[1]{\hspace{-1 ex} \nocite{#1}\citenum{#1}}
 \mciteErrorOnUnknownfalse
\definecolor{Gray}{gray}{0.9}
\author{Stefan Heinen}
\author{Max Schwilk}
\author{Guido Falk von Rudorff}
\author{O. Anatole von Lilienfeld}
\affiliation{Institute of Physical Chemistry and National Center for Computational Design and Discovery of Novel Materials (MARVEL), Department of Chemistry, University of Basel, Klingelbergstrasse 80, CH-4056 Basel, Switzerland}

\email{anatole.vonlilienfeld@unibas.ch}

\title{Machine learning the computational cost of quantum chemistry}

\begin{document}

\makeatletter
\setlength\acs@tocentry@height{10.1cm}
\setlength\acs@tocentry@width{5.1cm}
\makeatother
\begin{tocentry}
\begin{center}
\includegraphics[width=0.99\textwidth]{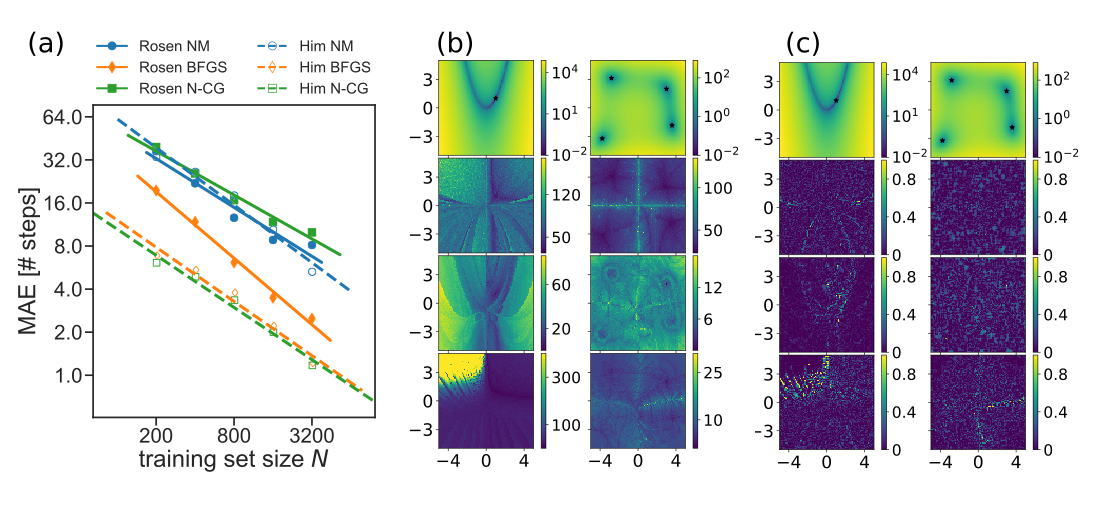}
\end{center}
\end{tocentry}
\onecolumn
\begin{abstract}
%Statistical learning, in particular kernel ridge regression, is applied to the prediction of the computational cost of various types of quantum chemistry calculations within a given chemical space and the results are used to address the scheduling problem on small, medium, and large size compute clusters.
Computational quantum mechanics based molecular and materials design campaigns consume increasingly more high-performance compute resources, making improved job scheduling efficiency desirable in order to reduce carbon footprint or wasteful spending.  
We introduce quantum machine learning (QML) models of the computational cost of common quantum chemistry tasks.  
For 2D non-linear toy systems, single point, geometry optimization, and transition state calculations the out of sample prediction error of QML models of wall times decays systematically with training set size. 
We present numerical evidence for a toy system containing two functions and three commonly used optimizer and for thousands of organic molecular systems including closed and open shell equilibrium structures,  as well as transition states. 
Levels of electronic structure theory considered include B3LYP/def2-TZVP, MP2/6-311G(d), local CCSD(T)/VTZ-F12, CASSCF/VDZ-F12, and MRCISD+Q-F12/VDZ-F12.
In comparison to conventional indiscriminate job treatment,
QML based wall time predictions significantly improve job scheduling efficiency for all tasks
after training on just thousands of molecules. 
Resulting reductions in CPU time overhead range from 10\% to 90\%.

%Given our results, we estimate potential world wide savings in the order of 5 $\cdot 10^{4}$ exaFLOP per second for quantum chemistry compute campaigns by adapting such QML models to optimize scheduling protocols. \\
\end{abstract}
%\tableofcontents
\twocolumn
%
%%%%%%%%%%%%%%%%%%%%%%%%%%%%%%%%%%%%%%%%%%%%%%%%%%%%%%%%%%%
% Introduction
%%%%%%%%%%%%%%%%%%%%%%%%%%%%%%%%%%%%%%%%%%%%%%%%%%%%%%%%%%%
\section{Introduction}
\label{sec:intro}
\begin{figure}[!ht]
    \centering
    \includegraphics[width=.49\textwidth]{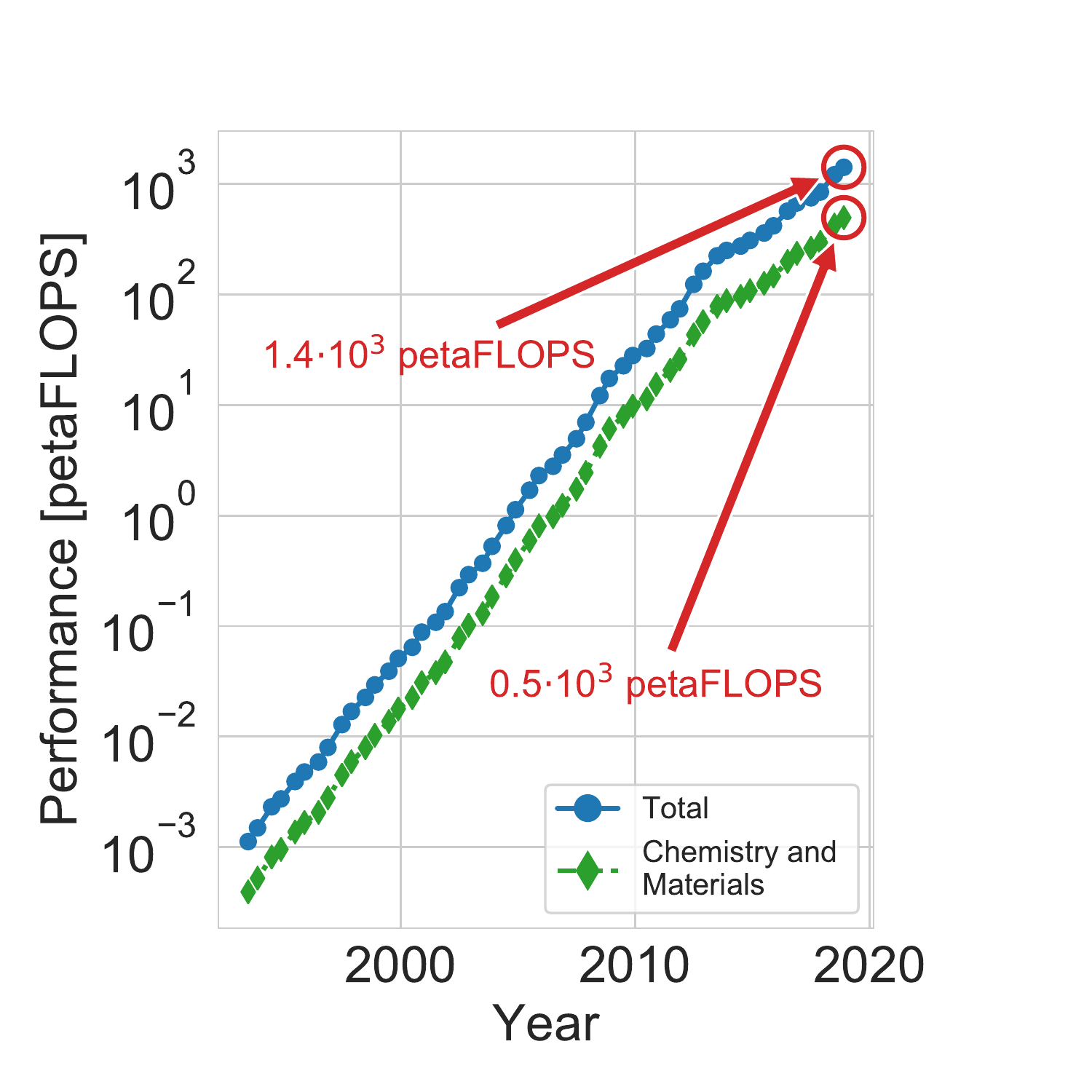}
    \caption{Compute resource growth of 500 fastest public supercomputers.\cite{top500} 
    Estimated use by chemistry and materials sciences corresponds to 35\%, corresponding to 2017 usage on Swiss National Supercomputing Center.\cite{CSCS}.}
    \label{fig:cscs_usage}
\end{figure}

% From SE -> Computers -> schedulers 
Solving Schr{\"o}dinger's equation, arguably one of the most important compute tasks for chemistry and materials sciences, with arbitrary accuracy is a NP hard problem\cite{Garey1990}.
This leads to the ubiquitous limitation that accurate quantum chemistry calculations typically suffer from computational costs scaling steeply and non-linearly with molecular size.
Therefore, even if Moore's law was to stay approximately valid~\cite{MooreLaw}, 
scarcity in compute hardware would remain a critical factor for the foreseeable future. 
%Global HPC resources for chemistry are substantial:
Correspondingly, chemistry and materials based compute projects have been consuming substantial CPU time at academic high-performance compute centers on national and local levels worldwide. 
For example, in 2017 research projects from chemistry and materials sciences used $\sim$25 and $\sim$35\% of the total available resources at Argonne Leadership Computing Facility~\cite{Argon} and at the Swiss National Supercomputing Center (CSCS)~\cite{CSCS}, respectively. 
In 2018, $\sim$30\% of the resources at the National Energy Research Scientific Computing Center~\cite{NERSC} were dedicated to chemistry and materials sciences and even $\sim$50\% of the resources of the ARCHER\cite{Archer} super computing facility over the past month (May 2019). 
Assuming a global share of $\sim$35\% for the usage of the Top 500 super computers (illustrated in Figure~\ref{fig:cscs_usage}) over the last 25 years, this would currently correspond to $\sim$0.5 exaFLOPS (floating point operations per seconds) per year.
But also on most of the local medium to large size university or research center compute clusters, atomistic simulation consumes a large fraction of available resources. 
For example, at sciCORE, the University of Basel's compute cluster, this fraction typically exceeds 50\%. 
Acquisition, usage, and maintenance of such infrastructures require substantial financial investments. Conversely, any improvements in the efficiency with which they are being used would result in immediate savings.
Therefore a lot of work is done to constantly improve hardware and software of HPCs, e.\ g. at the International Supercomputing Conference NVIDIA announced the support of the Advanced RISC Machines (Arm) CPUs, which allows to build extremely energy efficient exascale computers, by the end of the year\cite{NVIDIA}. 
Compute applications on such machines commonly rely on schedulers optimizing the simultaneous work load of thousands of calculations.
While these schedulers are highly optimized to reduce overhead, there is still potential for application domain specific improvements, mostly due to indiscriminate and humanly biased run time estimates specified by users. 
The latter is particularly problematic when it comes to ensemble set-ups characteristic for molecular and materials design compute campaigns with very heterogeneous compute needs of individual instances. 
One could use the scaling behaviour of methods to get sorted lists w.r.t wall times and improve scheduling by grouping the calculations by run time.
For example the bottleneck of a multi-configuration self-consistent field calculation (MCSCF) is in general the transformation of the Coulomb and exchange operator matrices into the new orbital basis during the macro-iterations.
This step scales as $n m^{4}$ with $n$ the number of occupied orbitals and $m$ the number of basis functions.
%The significant part of the MRCISD+Q-F12/VDZ-F12 run time was spent in the computation of the single-pair interactions, the pair-pair interactions, as well as the F12 contributions.
All Configuration Interaction Singles Doubles (CISD) schemes that are based on the Davidson algorithm\cite{Davidson1975JCoP} scale formally as $n^2 m^4$, where $n$ the number of correlated occupied orbitals and $m$ the number of basis functions.\cite{SherrillScalCI}
%This could improve scheduling for SP calculations.
As these methods (and basis sets) contain different scaling laws and geometry optimizations additionally depend on the initial geometry, a more sophisticated approach was applied: In this paper, we show how to use quantum machine learning (QML) to more accurately estimate run times in order to improve overall scheduling efficiency of quantum based ensemble compute campaigns.
Since the early 90's, an increasing number of research efforts from computer science has dealt with optimizing the execution of important standard classes of algorithms that occur in many scientific applications on HPC platforms,\cite{singh2007predicting,malakar2018benchmarking,GPUSched} but also with predicting memory consumption,\cite{rodrigues2016helping} or, more generally, the computational cost itself (see Refs.~\onlinecite{witt2019predictive,nemirovsky2018general} for two recent reviews).
Such predictive models may even comprise direct minimization of the estimated environmental impact of a calculation as the target quantity in the model.\cite{garg2011environment}
ML has already successfully been applied, however, towards improving scheduling itself,\cite{nemirovsky2017deep} or entire compute work flows.\cite{kousalya2017workflow,sahni2018cost}
Furthermore, a potentially valuable application in the context of quantum chemistry may be the run time optimization of a given tensor contraction scheme on a specific hardware by predictive modelling techniques\cite{liu2018using}.
Another noteworthy effort has been the successful run time modeling and optimization of a self-consistent field (SCF) algorithm on various computer architectures in 2011\cite{antony2011modelling} using a simple linear model depending on the number of retired instructions and cache misses.
% Hardware counters were obtained by the usage of instrumentation code and specific tools simulating cache behaviour.
%In this way, cache configuration could be slightly optimized for the studied self consistent field (SCF) application calculations.
Already in 1996, Papay et al. contributed a least square fit of parameters in graph based component-wise run time estimates in parallelized self consistent field computations of atoms.\cite{Papay}
Other noteworthy work in the field of computational chemistry is the prediction of the run time of a molecular dynamics code,\cite{mniszewski2015tadsim} or the prediction of the success of density functional theory (DFT) optimizations of transition metal species as a classification problem by Kulik and coworkers\cite{KulikFailPred}. 
%The prediction is used to analyze the dependency of the run time on the parameters of the program. % a task for which multiple executions of the program would be computationally too expensive.
In the context of quantum chemistry and quantum mechanical solid state computations, 
very little literature on the topic is found.
This may seem surprising, given the significant share of this domain on the overall HPC resource consumption (cf.\ Figure~\ref{fig:cscs_usage}).
To the best of our knowledge, there is no (Q)ML study that predicts the computational cost (wall time, CPU time, FLOP count) of a given quantum chemical method across chemical space.

%QML intro
Today, a large number of QML models relevant to quantum chemistry applications throughout chemical space exists~\cite{anatole-ijqc2013, QMLessayAnatole, rupp2018guest}.
Common regressors include Kernel Ridge Regression\cite{RuppPRL2012,AssessmentMLJCTC2013,SingleKernel2015,Bing2016,RaghusReview2016,googlePaper2017} (KRR), Gaussian Process Regression\cite{GPR} (GPR), or Artificial Neural Networks\cite{Montavon2013,ANI_IsayevRoitberg2017,schutt2017quantum,googlePaper2017,schnett,ounke} (ANN).
For the purpose of estimating run times of new molecules, and contrary to pure computer science approaches, we use the same molecular representations (derived solely from molecular atomic configurations and compositions) in our QML models as for modeling quantum properties.
As such, we view computational cost as a molecular ``quasi-property'' that can be inferred for new, out-of-sample input molecules, in complete analogy to other quantum properties, such as the atomization energy or the dipole moment.

In general, a quantum chemistry SCF calculation optimizes the parameters of a molecular wave function with a clear minimum in the self-consistent system of non-linear equations. I.\ e., the computational cost of a single point quantum chemistry calculation should be a reasonably smooth property over the chemical space. % because it scales with the number of electrons.
Pathological cases of SCF convergence failure are normally avoided by the careful choice of the quantum chemistry method for the single point (SP) calculation of a given chemical system.
For geometry optimization (GO) and transition state (TS) searches on the other hand  it is much harder to control the convergence, as a multitude of local minima and saddle points may exist on the potential energy surface defined by the degrees of freedom of the atomic coordinates in the molecule. 

We therefore first investigated the performance of ML models to learn the number of discrete steps of common optimizers applied to the minimum search of non-linear 2D functions that are known to cause convergence problems for many standard optimizers.
In a second step, we investigated the capabilities of QML to learn the computational cost for a representative set of quantum chemistry tasks, including  SP, GO, and TS calculations.
To provide numerical evidence for hardware independence of the cost of quantum chemistry calculations, we trained a model on FLOPS as a ``clean'' measurement. 

% This ambition requires to train a model for each type of calculation  (e.g. geometry optimizations or single point calculations)  to learn core hours on different molecules.
% This would assume to treat the calculation cost (CPU or wall times) as a molecular property.
% The key question is whether molecular structure based representations provide an adequate feature space such that learning is possible based on the given training data.\\\\
%
%%%%%%%%%%%%%%%%%%%%%%%%%%%%%%%%%%%%%%%%%%%%%%%%%%%%%%%%%%%
% Methods
%%%%%%%%%%%%%%%%%%%%%%%%%%%%%%%%%%%%%%%%%%%%%%%%%%%%%%%%%%%
\section{Data}
\label{sec:Data}
%%%%%%%%%%%%%%%%%%%%%%%%%%%%%%%%%%%%%%%%%%%%%%%%%%%%%%%%%%%
% Data Sets
%%%%%%%%%%%%%%%%%%%%%%%%%%%%%%%%%%%%%%%%%%%%%%%%%%%%%%%%%%%

% data sets, state of the art
All QML approaches rely on large training data sets.
Comprehensive subsets of the chemical space 
of closed shell organic molecules %close to their equilibrium structure
have been created in the past.
%, e.\ g., the QM9\cite{DataPaper2014} data set which is derived from a subset of the GDB17\cite{GDB17}. % set of Simplified Molecular Input Line Entry System (SMILES) strings.\cite{w1988,www1989}
The QM9\cite{DataPaper2014} data set of DFT optimized 3D molecular structures was derived from the GDB17\cite{GDB17} data set of Simplified Molecular Input Line Entry System (SMILES) strings\cite{w1988,www1989}.
This data set contains drug like molecules of broad scientific interest. 
GDB17 is an attempt to systematically generate molecules as mathematical graphs based on rules of medicinal chemistry, removing the bias of pre-existing building blocks in structure selection. 
QM9 itself is a well established benchmark data set for quantum machine learning where many different ML models were tested on\cite{RuppPRL2012,BobPaper,Bing2016,HDAD,DTNN,google,SOAP, FCHL,schnett,Unke_NN,Lubbers_2018,wavelet} and also contains many molecules which are commercially available and reported on many chemical data bases.
Further relevant data sets in the literature include, among others, reaction networks,\cite{reiherRxnData2018} closed shell ground state organometallic compounds,\cite{volcaMeetsML2018} or non-equilibrium structures of small closed shell organic molecules.\cite{smith2017ani}
Yet, regions of chemical space that may involve more sophisticated and costly quantum chemistry methods, such as open shell and strongly correlated systems\cite{C7SC01247K,porph} or chemical reaction paths, are still strongly underrepresented. %TODO cite max if not absent, in literature.% data sets, in this study
For this study, we first generated two toy systems of non-linear functions known to be difficult for many standard optimization methods.
We used KRR to predict the number of optimization steps needed to find the functions' closest minimum for a systematically chosen set of starting points.
The test case of optimizing analytical functions explores the fundamental question of learning computational cost of a non-linear optimization problem outside the added complexity of quantum chemistry calculations.
We then have generated measures of the computational cost associated to seven tasks which reflect variances of three common use cases: single point (SP), geometry optimization (GO) and transition state (TS) search calculations.
\subsection{Toy System}
 To demonstrate that it is possible to learn the number of steps of an optimization algorithm, we apply our machine learning method to two cases from function optimization theory:
 quantifying the number of steps for an optimizer. The functions in question are the Rosenbrock function\cite{Rosenbrock}
\begin{equation}
    f(x,y) = (1 - x)^2 + 100(y - x^2)^2
\end{equation} 
and the Himmelblau function\cite{Himmelblau}
\begin{equation}
    f(x,y) = (x^2 + y - 11)^2 + (x + y^2 -7)^2
\end{equation} 
The fucntions are shown in the top row of Figure \ref{fig:tool_kit} b) and c).
We applied three representative optimizers in their SciPy 1.3.1\cite{ScyPy} implementation on both functions: the ``NM'' simplex algorithm (Nelder-Mead\cite{Nelder}), the gradient based ``BFGS'' algorithm\cite{BFGS}, and an algorithm using gradients and hessians (Conjugate Gradient with Newton search ``N-CG''\cite{NewtonCG}).
For every function and optimizer we performed $10200$ optimizations from different starting points on a cartesian grid over the domain $-5 \le x, y \le 5$ in steps of $0.1$.
The minimum of the Rosenbrock function and the four minima of the Himmelblau function lie within this domain.
Figure \ref{fig:tool_kit} b) row two, three, and four show a heatmap of the number of optimization steps for NM, BFGS, and N-CG, respectively, for Rosenbrock (left column) and Himmelblau (right column).
Generally, the minimum searches on the Himmelblau function required much fewer steps (mostly reached after a few tens of iterations).
While the gradient based optimizer BFGS clearly outperforms NM for both functions, the N-CG optimization of the Rosenbrock function did not converge with a iteration limit of 400 for a set of points in the region of $x < -0.5$  and $y > 2.5$. 
A very small step size for the N-CG algorithm implementation in SciPy in the critical region is responsible for the slow convergence.

\subsection{Quantum Data Sets}
We have considered coordinates coming from three different data sets (QM9, QMspin, QMrxn) corresponding to five levels of theory %electron correlation treatments
(CCSD(T), MRCI, B3LYP, MP2, CASSCF) and four basis set sizes. 
Molecules in the three different data sets consist of the following:
% the 3 data sets (in detail)
\begin{enumerate}[i)]
\item QM9 contains 134k small organic molecules in the ground state local minima with up to nine heavy atoms which are composed of H, C, N, O, and F. 
All coordinates were published in 2014~\cite{DataPaper2014}. Here, we also report the relevant timings.
\item QMspin consists of carbenes derived from QM9 molecules containing calculations of the singlet and triplet state, respectively, with a state-averaged CASSCF(2e,2o) reference wave function (singlet and triplet ground states with equal weights). 
The entirety of this data set will be published elsewhere, here we only provide timings and QM9 labels. 
\item QMrxn consists of reactants and S$_N$2 transition states of small organic molecules with a scaffold of C$_{2}$H$_{6}$ which was functionalized with the following substituents: \mbox{-NO$_{2}$}, -CN, -CH$_{3}$, -NH$_{2}$, -F, -Cl and -Br.
The entirety of this data set will be published elsewhere, here we only provide timings and geometries. 
\end{enumerate}

\subsection{Quantum Chemistry Tasks}
\begin{figure}[!ht]
    \centering
    \includegraphics[width=.49\textwidth]{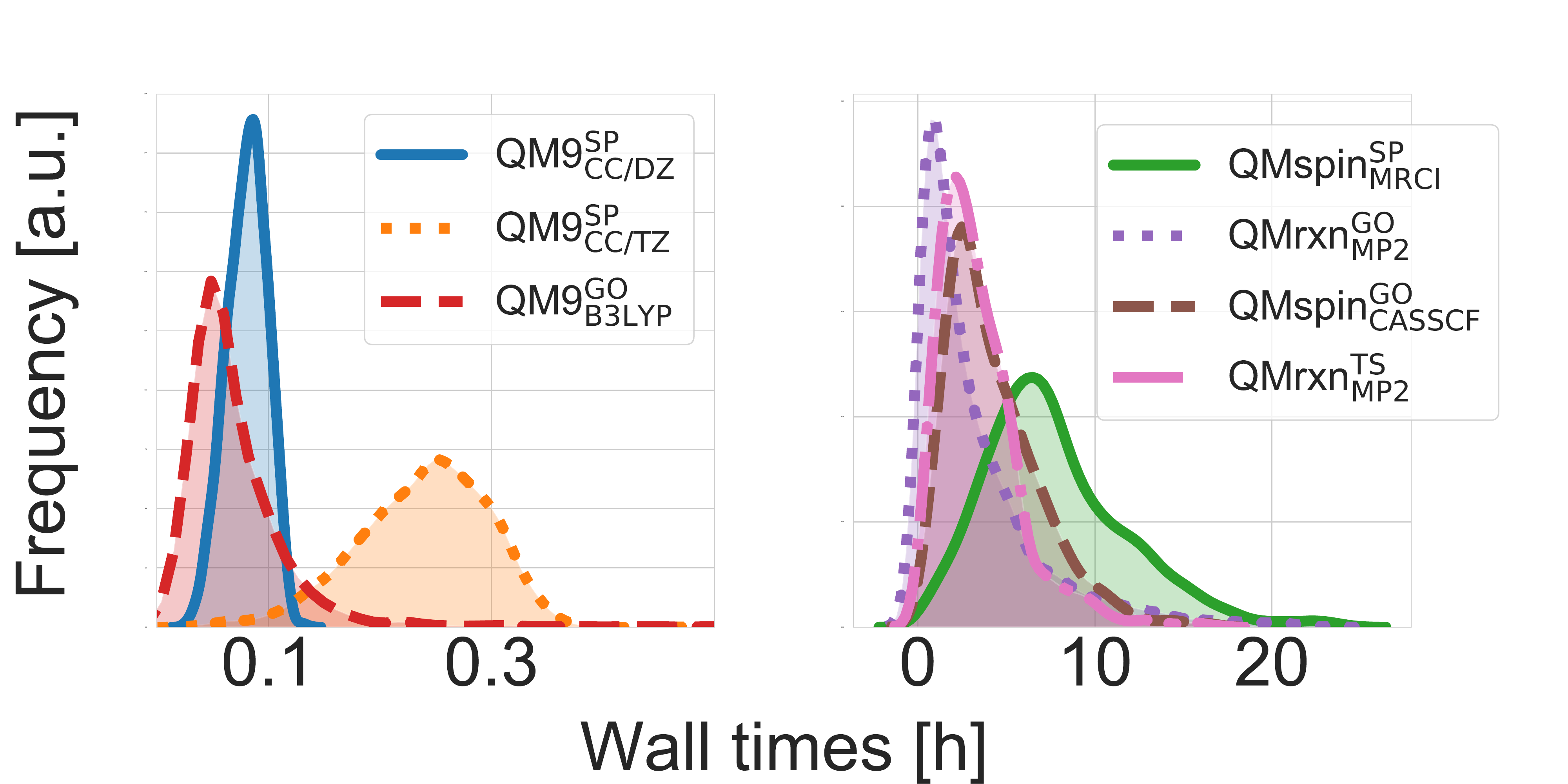}
    \caption{Wall time distribution of all tasks using kernel density estimation.}
    \label{fig:dens}
\end{figure}

% the 7 tasks
The three data sets were then divided into the seven following tasks for which timings were obtianed (See also Table \ref{tab:datasets}):
% the seven tasks (with methods and basis sets)
\begin{description}
\item[QM9$^{\textrm{SP}}_{\textrm{CC/DZ}}$]
\label{subsec:ccdz}
5736 PNO-LCCSD(T)-F12/VDZ-F12\cite{Schwilk2017JCTC,Ma2017JCTC_1,Ma2017JCTC_2} single point energy timings.
%Details on the coupled cluster calculations will be published in a separate publication.\cite{Schwilk2019tbp}
Details of 
%on this data set of strongly correlated organic molecules and 
the calculation results other than timings are subject of a separate publication.\cite{Schwilk2019tbp}
\item[QM9$^{\textrm{SP}}_{\textrm{CC/TZ}}$]
3497 PNO-LCCSD(T)-F12/VTZ-F12 single point energy timings.
\item[QMspin$^{\textrm{SP}}_{\textrm{MRCI}}$]
2732 single point calculations using MRCISD+Q-F12/VDZ-F12\cite{Knowles1988CPL,Werner1988JCP,Shiozaki2011JCP,Shiozaki2013MP}.
Details of 
%on this data set of strongly correlated organic molecules and 
the calculation results other than timings are subject of a separate publication.\cite{Tahchieva2019tbp}
\item[QM9$^{\textrm{GO}}_{\textrm{B3LYP}}$]
3724 geometry optimization timings with initial B3LYP/6-31G*\cite{B3LYP_functional,LYP} geometries optimizing at the B3LYP/def2-TZVP level of theory.
\item[QMrxn$^{\textrm{GO}}_{\textrm{MP2}}$]
\label{subsec:qmrxn}
8148 geometry optimization timings on MP2/6-311G(d) level of theory.
\item[QMspin$^{\textrm{GO}}_{\textrm{CASSCF}}$]
1595 CASSCF(2e,2o)[Singlet]/VDZ-F12\cite{Werner1985JCP,Busch1991JCP} geometry optimization timings.
\item[QMrxn$^{\textrm{TS}}_{\textrm{MP2}}$]
1561 timings of transition state searches on MP2 level of theory.
\end{description}
Further details on the data sets can be found in section 1 of the supporting information (SI).
A distribution of the properties (wall times) of the seven tasks is illustrated in Figure \ref{fig:dens}.
Single point calculations (the two \textbf{QM9$^{\textrm{SP}}_{\textrm{CC}}$} tasks) and the geometry optimization (task \textbf{QM9$^{\textrm{GO}}_{\textrm{B3LYP}}$}) have wall times smaller than half an hour.
In general, the smaller the variance in the data, the less complex the problem and the easier it is for the model to learn the wall times.
For geometry optimizations and more exact (also more expensive) methods (task \textbf{QMspin$^{\textrm{SP}}_{\textrm{MRCI}}$} and \textbf{QMspin$^{\textrm{GO}}_{\textrm{CASSCF}}$}) the average run time is $\sim$ 9 hours.
With a larger variance in the data the problem is more complex (higher dimensional) and the learning is more difficult (higher off-set).
%three common types of systems: (i) $\sim$1'500 reactive organic molecules (transition state, reactant, product) (\textbf{QMrxn}), (ii) $\sim$ small organic closed shell molecules (\textbf{QM9}), and (iii) strongly correlated carbenes (\textbf{QMspin}).

%These three molecular data sets were used within seven data sets corresponding to different calculation tasks, including single point (SP), geometry optimizations (GO), and transition state search (TS) calculations.
%The method levels range from semi-empirical methods (PM6) over density functional theory (DFT), M{\o}ller-Plesset perturbation theory (MP2), and Coupled Cluster (CC) to complete active space SCF (CASSCF) and multi-reference configuration interaction (MRCI).
%Furthermore, we covered basis sets of different sizes and characteristics, namely representatives of the split valence basis sets of Pople and coworkers\cite{binkley1980self,Petersson1988,Petersson1991} [here 6-311G(d)], the Karlsruhe basis sets\cite{Weigend2005,Weigend2006} popular for DFT (here def2-TZVP), and finally the correlation consistent F12 basis sets of Peterson \emph{et al.}\cite{Kirk:VnZF1208} (here V$n$Z-F12, $n$=\{D, T\}).
%The seven use case data sets are outlined in Table~\ref{tab:datasets}.

%
\begin{table*}
\scalebox{0.8}{
\begin{tabular}{l | c c c | c c c | c}
    \rowcolor{Gray}
    \textbf{Task}    & 
    \textbf{QM9$^{\textrm{SP}}_{\textrm{CC/DZ}}$} & \textbf{QM9$^{\textrm{SP}}_{\textrm{CC/TZ}}$} & \textbf{QMspin$^{\textrm{SP}}_{\textrm{MRCI}}$} & \textbf{QM9$^{\textrm{GO}}_{\textrm{B3LYP}}$} & \textbf{QMrxn$^{\textrm{GO}}_{\textrm{MP2}}$} & \textbf{QMspin$^{\textrm{GO}}_{\textrm{CASSCF}}$} & \textbf{QMrxn$^{\textrm{TS}}_{\textrm{MP2}}$}\\
    \textbf{Use case}     & \multicolumn{3}{c|}{ SP } & \multicolumn{3}{c|}{ GO } & TS  \\
    \rowcolor{Gray}
    \textbf{Data set}  & \multicolumn{2}{c}{ QM9 } & QMspin & QM9 & QMrxn & QMspin  & QMrxn \\
    \textbf{Level}    &  CCSD(T) & CCSD(T) & MRCI & B3LYP& MP2 &  CASSCF  & MP2 \\
    \rowcolor{Gray}
    \textbf{Basis set}    &  
    VDZ-F12\cite{Kirk:VnZF1208}     &
    VTZ-F12\cite{Kirk:VnZF1208}     & 
    VDZ-F12\cite{Kirk:VnZF1208}     & 
    def2-TZVP \cite{Weigend2005,Weigend2006} & 6-311G(d)\cite{binkley1980self,Petersson1988,Petersson1991} &
    VDZ-F12\cite{Kirk:VnZF1208}  &
    6-311G(d)\cite{binkley1980self,Petersson1988,Petersson1991} \\
    \textbf{Size} & 5736 & 3497 & 2732 & 3724 & 8148 & 1595 & 1561 \\ 
    \rowcolor{Gray}
    \textbf{Code} & Molpro & Molpro & Molpro & Molpro & ORCA & Molpro & ORCA \\
%
%    \textbf{$\bar{N}$} & 8.59 & 8.79 & 8.36 & 6.99 & 8.26 & 7.91 & 7.39 \\ 
%
%    \textbf{$\sigma_{\textrm{BoB}}$} & 819.2 & 204.8 & 51.2 & 51.2 & 409.6 & 51.2 & 409.6 \\
%%
%    \rowcolor{Gray}
%    \textbf{$\sigma_{\textrm{FCHL}}$} & 12.8 & 25.6 & 25.6 & 409.6 & 25.6 & 204.8 & 204.8 \\
%%    

%    \textbf{Label}     & \textbf{A} & \textbf{B} & \textbf{C} & \textbf{D} & \textbf{E} & \textbf{F} & \textbf{G}   \\
%    \hline \hline
%    \textbf{Method}    & \makecell{MP2/\!/PM6} & MP2 & CC & CC & \makecell{B3LYP} & MRCI   & \makecell{CASSCF(2e,2o)(S=0)/\!/\\ROKS-B3LYP(S=1)}\\
%    %
%    \textbf{Basis Set} & \makecell{6-311G(d)} & 6-311G(d) & VDZ-F12 & VTZ-F12 & \makecell{TZVP/\!/6-31G*} & --   & \makecell{VDZ-F12/\!/def2-TZVP}\\
%    %
%    \textbf{Type} & GO & TS & SP       & SP        & GO & SP        & GO\\
%    \textbf{Data Set}   & Reactions & Reactions & Closed shell & Closed shell & Closed shell & Carbenes & Carbenes \\
%    \textbf{Data Set Size} & 8148      &1561       &   9843    &   3497    &   5001    &   3562    &   2099\\
%    $\bar{N}$ & 8.26     &  7.39    &   8.59    &  8.79      &  6.99      &   8.36    &   7.91    \\
%    %
%    \hline

    %    $\lambda$     & 10$^{-7}$      &   10$^{-7}$    &   10$^{-7}$    &   10$^{-7}$    &   10$^{-7}$    &   10$^{-7}$    &   10$^{-7}$\\
\end{tabular}}
\caption{Seven tasks used in this work generated from three data sets (QM9, QMspin, QMrxn), using three use cases (SP, GO, TS) on different levels of theory and basis sets.}
%Machine learning model hyperparameters $\sigma$ and $\lambda$ are reported as well. }
\label{tab:datasets}
\end{table*}
%}
%\end{figure*}

\subsection{Timings, Code, and Hardware}
% calculations. ele struct code
The calculations were run on three compute clusters, namely our in-house compute cluster, the Basel University cluster (sciCORE) and the Swiss national supercomputer Piz Daint at CSCS.
We used two electronic structure codes to generate timings.
Molpro\cite{Molpro} was used to extract both CPU and wall times for data sets i) and ii),
and ORCA\cite{orca} was used to extract wall times for data set iii).
%To generate timings, we used two electronic structure codes: ORCA\cite{orca} for data set iii) where we extracted wall times and Molpro\cite{Molpro} for data set i) and ii) where we extracted both, CPU and wall times.
Further information of the data sets, the hardware, and the calculations can be found in section 3 to 4 of the SI. %\ref{SI})
%In section 1 of the supporting information (SI) further details on hyperparameter, used hardware, number of cores and versions of the used  quantum chemical software packages are given.

%For data sets i) and iii) coordinates and timings can be found in the supporting information (SI).
%For data set ii), the index of the QM9 molecule from which the carbene molecules have been derived together with the timings are given in the SI.
%and represent a range of typical calculations common in the field of computational quantum chemistry.
%\subsection{FLOPS}
The retired floating point operations (FLOP) count of the local coupled cluster calculation task \textbf{QM9$^{\textrm{SP}}_{\textrm{CC/DZ}}$} was obtained as follows:
\label{subsec:flop}
The number of FLOPs have been computed with the \emph{perf} Linux kernel profiling tool\cite{PerfVersion} for data set \textbf{QM9$^{\textrm{SP}}_{\textrm{CC/DZ}}$}.
\emph{perf} allows profiling of the kernel and user code at run time with little CPU overhead and can give FLOP counts with reasonable accuracy. 
FLOP count is an adequate measure of the computational cost when the program execution is CPU bound by numerical operations, which is given in the PNO-LCCSD(T)-F12 implementation\cite{Schwilk2017JCTC,Ma2017JCTC_1,Ma2017JCTC_2,wires_cc} in Molpro. %such as tensor calculus.
\section{Methods}
\label{subsec:ml}
\subsection{Quantum Machine Learning}
In this study, we used kernel based machine learning methods which were initially developed in the 1950s\cite{kriging} and belong to the supervised learning techniques.
In ridge regression, the input is mapped into a feature space and fitting is applied there.
However, the best feature space is \emph{a priori} unknown, and its construction is computationally hard.
The ``kernel trick'' offers a solution to this problem by applying a kernel $k$ on a representation space $\mathcal{R}$ that yields inner products of an implicit  high dimensional feature space:
the Gram matrix elements $k(\mathbf{x}_i, \mathbf{x}_j)$ of two representations $\mathbf{x} \in \mathcal{R}$ between two input molecules $i$ and $j$ are the inner products $\langle i , j\rangle$ in the feature space. For example, 
\begin{equation}
     k(\mathbf{x}_i, \mathbf{x}_j) = \mathrm{exp}\left( - \frac{|| \mathbf{x}_i - \mathbf{x}_j ||_1}{\sigma} \right)
     \label{eq:k_laplace}
\end{equation}
or
\begin{equation}
     k(\mathbf{x}_i, \mathbf{x}_j) = \mathrm{exp}\left( - \frac{|| \mathbf{x}_i - \mathbf{x}_j ||^{2}_{2}}{2\sigma^{2}} \right)
     \label{eq:k_gauss}
\end{equation}
with  $\sigma$ as the length scale hyperparameter, represent commonly made kernel choices, the Laplacian (eq.\ \ref{eq:k_laplace}) or Gaussian kernel (eq.\ \ref{eq:k_gauss}).
Fitting coefficients $\pmb{\alpha}$ can then be computed in input space via the inverse of the kernel matrix $[\mathbf{K}]_{ij} = k(\mathbf{x}_i, \mathbf{x}_j)$: 
\begin{equation}
    \pmb{\alpha} = (\mathbf{K} + \lambda \mathbf{I})^{-1} {\bf y}
    \label{eq:train}
\end{equation}
where $\lambda$ is the regularization strength, typically very small for calculated noise-free quantum chemistry data. 

Hence, kernel ridge regression (KRR) learns a mapping function from the inputs $\mathbf{x}_i$, in this case the representation of the molecule, to a property $y^{\rm est}_q(\mathbf{x}_q)$, given a training set of $N$ reference pairs $\{ (\mathbf{x}_i, y_i) \}^{N}_{i=1}$.
Learning in this context means interpolation between data points of reference data $\{ (\mathbf{x}_i, y_i) \}$ and target data $\{ (\mathbf{x}_q, y^{\mathrm{est}}_q) \}$.
A new property $y^{\mathrm{est}}_q$ can then be predicted via the fitting coefficients and the kernel:% (instead of a design matrix and fitting coefficients in a high dimensional feature space):
\begin{equation}
    y^{\rm est}_q(\mathbf{x}_q) = \sum_{i}^{N} \alpha_i \cdot k(\mathbf{x}_i, \mathbf{x}_q)
    \label{eq:pred}
\end{equation}

%To make use of KRR we relied on two representations in this study.
For the toy systems, a Laplacian kernel was used, the representation corresponding simply to the starting point ($\mathbf{x} = (x,y)$) of the optimization runs.
For the purpose of learning of the run times, we used two widely used representations, namely Bag of Bonds (BoB)\cite{BobPaper} with a Laplacian kernel. 
BoB is a vectorized version of the Coulomb Matrix (CM)\cite{RuppPRL2012} that takes the Coulomb repulsion terms for all atom to atom distances and packs them into bins, scaled by the product of the nuclear charges of the corresponding atoms.
%However, BoB does not only contain geometry information (distances) but also weights the individual bins of these distances with the corresponding nuclear charges of the atoms.
This representation does not provide a strictly unique mapping~\cite{FourierDescriptor, Bing2016} which may deteriorate learning in some cases (\textit{vide infra}).
The second representation used was atomic FCHL~\cite{FCHL} with a Gaussian kernel.
FCHL accounts for one-, two-, and three-body terms (whereas BoB only contains two-body terms).
The one-body term encodes group and period of the atom, the two-body term contains interatomic distances $R$, scaled by $R^{-4}$, and the three-body terms in addition contain angles between all atom triplets scaled by $R^{-2}$. 

To determine the hyperparameters $\sigma$ and $\lambda$, the reference data was split into two parts, the training and the test set.
The hyperparameters were optimized only within the training set using random sub-sampling cross validation.
%For each pair ($\sigma$, $\lambda$) a KRR model was trained and the predictions were tested on a hold out set, which was not part of the training set.
%The optimal combination of hyperparameters result in the smallest MAE.
%Once the hyperparameters have been optimized, the model was verified against the test set.
To quantify the performance of our model, the test errors, measured as mean absolute errors (MAE), were calculated as a function of training set size.
The leading error term is known to be inversely proportional to the amount of training points used:\cite{StatError_Muller1996}
\begin{equation}
    \mathrm{MAE} \approx a / N^{b}
    \label{eq:error}
\end{equation}
The learning curves should then result in a decreasing linear curve with slope $b$ and offset $\log a$:
\begin{equation}
    \log(\mathrm{MAE}) \approx \log(a) - b\log(N)
    \label{eq:logerror}
\end{equation}
where $a$ is the target similarity which gives an estimate of how well the mapping function describes the system\cite{Bing2016} and $b$ is the slope being an indicator for the effective dimensionality\cite{amons2017}.
Therefore, good QML models are linearly decaying, have a low offset $\textrm{log}(a)$ (achieved by using more adequate representations and/or base-line models~\cite{DeltaPaper2015}),
and have steep slopes (large $b$).\\\\
%
%For the data set A we also tried to improve our representation by adding the forces per atom of the first step of the geometry optimization to the BoB vector (from now on called \textit{force} BoB or data set A') weighted by an optimized hyperparameter $\gamma$ to scale both vectors (BoB and force vectors) to the same dimensions. The forces should give additional information to the representation of how close the molecule is to the relaxed structure and this should, as discussed in the previous paragraph, lower the off set log($a$) (target similarity) of the learning curve.\\\\
%
%\subsubsection{Setup}
%\label{subsubsec:setup}
For each task, QML models of wall times were trained and subsequently tested on out-of-sample test set which was not part of the training. 
As input for the representations the initial geometries of the calculations were used. 
%{\color{red}For every task the same procedure was used to make it more transferable.}
%%%%%%%%%%%%%%%
% from results
%%%%%%%%%%%%%%%
%In general, it is  more difficult to learn timings of geometry optimizations or transition state searches than timings of single point calculations.
%Similar challenges were addressed by Kulik and coworkers\cite{KulikFailPred} when predicting the successful outcome of DFT geometry optimizations of transition metal species as a classification problem.
%When using support vector machines and ANN with connectivity-only representations as input they were able to optimize their job list by rather reliably discarding jobs that will not properly converge for the given method.
%They could further improve their results by using representations containing electronic and geometric features obtained during the geometry optimization.
To improve the predictions of geometry optimizations for the task \textbf{QMspin$^{\textrm{GO}}_{\textrm{CASSCF}}$}, we split the individual optimization steps into the first step (GO1) and the subsequent steps (GO2), because the first step takes on average $\sim$20\% more time than the following steps (for more details we refer to section 1.4 of the SI).
%For the task \textbf{QMspin$^{\textrm{GO}}_{\textrm{CASSCF}}$}, we additionally analyzed the individual optimization steps, namely the first step (\textbf{GO1} subset) and the subsequent steps (\textbf{GO2} subset).
For learning the timings of the geometry optimization task GO2, we took the geometries obtained after the first optimization step.
%{\color{red}Also for the learning of the GO steps the procedure was not changed.}

\begin{figure}[!ht]
    \centering
    \includegraphics[width=.49\textwidth]{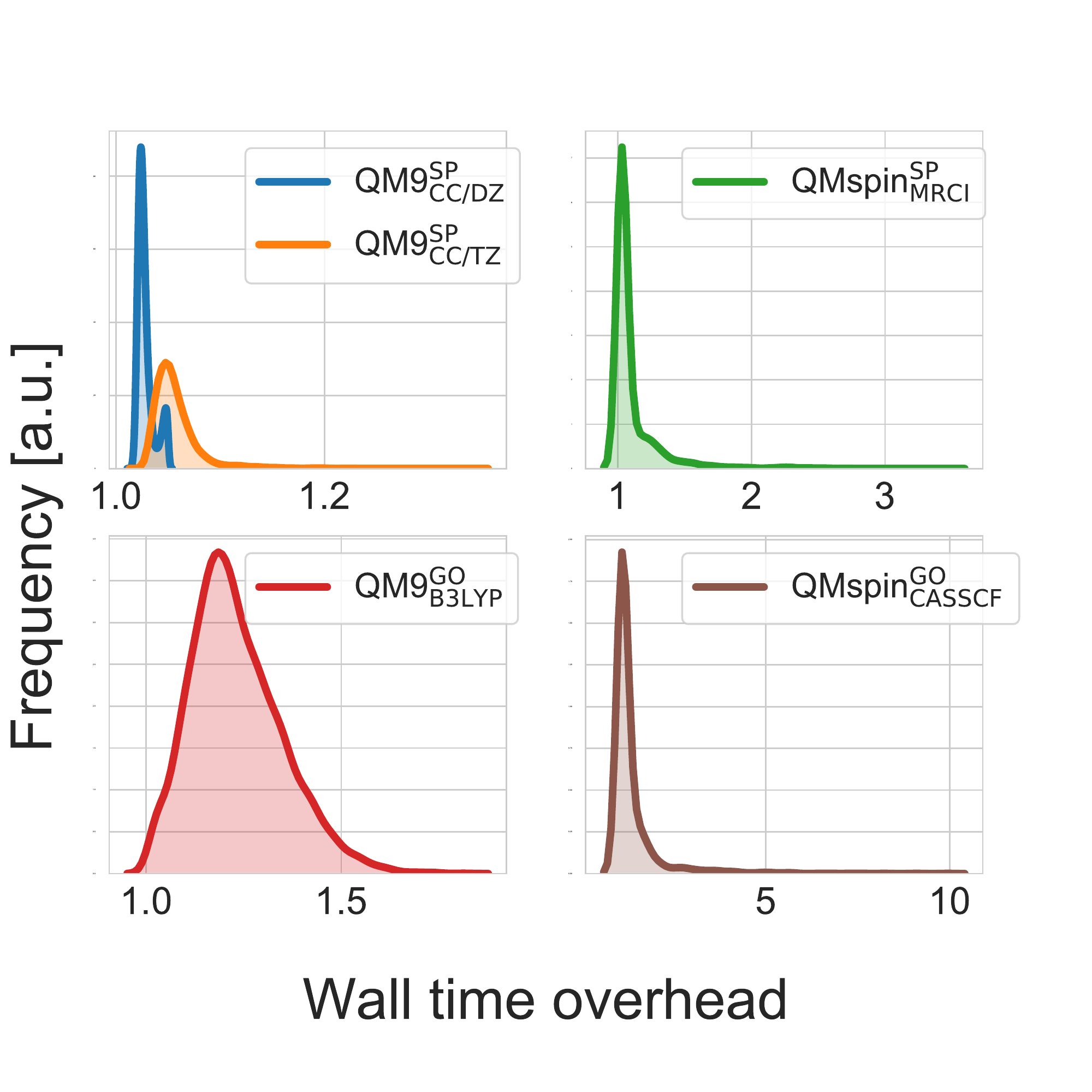}
    \caption{Wall to CPU time ratio (using kernel density estimation) for Molpro calculations to identify runs with high wall time overhead due to heavy I/O load on clusters.}
    \label{fig:overhead}
\end{figure}

As input for the properties, wall times were normalized with respect to the number of electrons in the molecules.
Figure \ref{fig:overhead} shows the wall time overhead (CPU time to wall time ratio) for calculations run with Molpro. To remove runs affected by heavy I/O, wall time overheads higher than 3\%, 5\%, 10\%, 30\%, and 50\% were excluded from the tasks \textbf{QM9$^{\textrm{SP}}_{\textrm{CC/DZ}}$}, \textbf{QM9$^{\textrm{SP}}_{\textrm{CC/TZ}}$}, \textbf{QMspin$^{\textrm{SP}}_{\textrm{MRCI}}$}, \textbf{QMspin$^{\textrm{GO}}_{\textrm{CASSCF}}$}, and \textbf{QM9$^{\textrm{GO}}_{\textrm{B3LYP}}$}, respectively. 
%Normalization of the timings that take into account the formal scaling of the method were also tested.
In order to generate learning curves for all the seven tasks, all timings were normalized with respect to the median of the test set to get comparable normalized mean absolute errors (MAE). 
%so that we are able to compare the different time scales of the different types of calculations.
The resulting wall time out-of-sample predictions were used as input for the scheduling algorithm.
Whenever the QML model predicted negative wall times, the predictions were replaced by the median of all non-negative predictions. \\
All QML calculations have been carried out with {QMLcode}\cite{qmlcode2017}.
Wall times and CPU times (Molpro) and wall times (ORCA) for all the seven tasks, as well as QML scripts can be found in the SI.
%
%%%%%%%%%%%%%%%%%%%%%%%%%%%%%%%%%%%%%%%%%%%%%%%%%%%%%%%%%%%
% Scheduling
%%%%%%%%%%%%%%%%%%%%%%%%%%%%%%%%%%%%%%%%%%%%%%%%%%%%%%%%%%%
\subsection{Application: Optimal Scheduling}
% TODO SI
\subsubsection{Job Array and Job Steps}
In many cases, efforts in computational chemistry or materials design require the evaluation of identical tasks on different molecules or materials.
Distributing those tasks across a compute cluster is typically done in one of two ways.
When using job arrays, the scheduler assigns compute resources to each calculation separately, such that the individual calculation is queued independently.
This approach typically extends the total wall time, and has little overhead with the jobs themselves but leads to inefficiencies for the scheduler since the individual wall time estimate of each job needs to be (close to) the maximum job duration.

In the second approach, there are only few jobs submitted to the scheduler and tasks are executed in parallel as job steps.
The first approach has little overhead with the jobs themselves but can lead to inefficiencies.
The second approach yields inefficiencies due to lack of load balancing. 
These two common methods require no knowledge of the individual run time of each task, and usually rely on a conservative run time estimate in practice.

\subsubsection{Scheduling Simulator}
Using the QML based estimated absolute timings turns the scheduling of the remaining calculations into a bin packing problem.
For this problem we used the heuristic first fit decreasing (FFD) algorithm which takes all run time estimates for all tasks, sorts them in decreasing order and chooses the longest task that fits into the remaining time of a compute job (for more details on FFD, see section 2 in the SI).
If there is no task left that is estimated to fit into a gap, then no task is chosen and resources are released early.

We implemented a job scheduling simulator assuming idempotent uninterruptible tasks for all three job schedulers: Conventional job arrays, conventional job steps, and our new QML based job scheduler.
%This allowed for estimating the performance of all three approaches without having to actually perform all quantum chemistry calculations over and over again.
Using a simulator is particularly useful because the duration of the job array and job step approaches depend on the (random) order of the jobs, and therefore requires averaging over multiple runs. 
We used this simulator in the context of two environments: our university cluster sciCORE (denoted \textit{S}) where users are allowed to submit single-core jobs and the Swiss national supercomputer (CSCS, denoted \textit{L}) where users are only allowed to allocate entire compute nodes of 12 cores.
In all cases, we assumed that starting a new job via the scheduler takes 30 seconds and that every job queues for one hour. These numbers have been observed for queuing statistics of sciCORE and CSCS.
%
%%%%%%%%%%%%%%%%%%%%%%%%%%%%%%%%%%%%%%%%%%%%%%%%%%%%%%%%%%%
% Results
%%%%%%%%%%%%%%%%%%%%%%%%%%%%%%%%%%%%%%%%%%%%%%%%%%%%%%%%%%%
\section{Results and Discussion}
\subsection{Toy System}
\begin{figure*}[!ht]
    \centering
    \includegraphics[width=.9\textwidth]{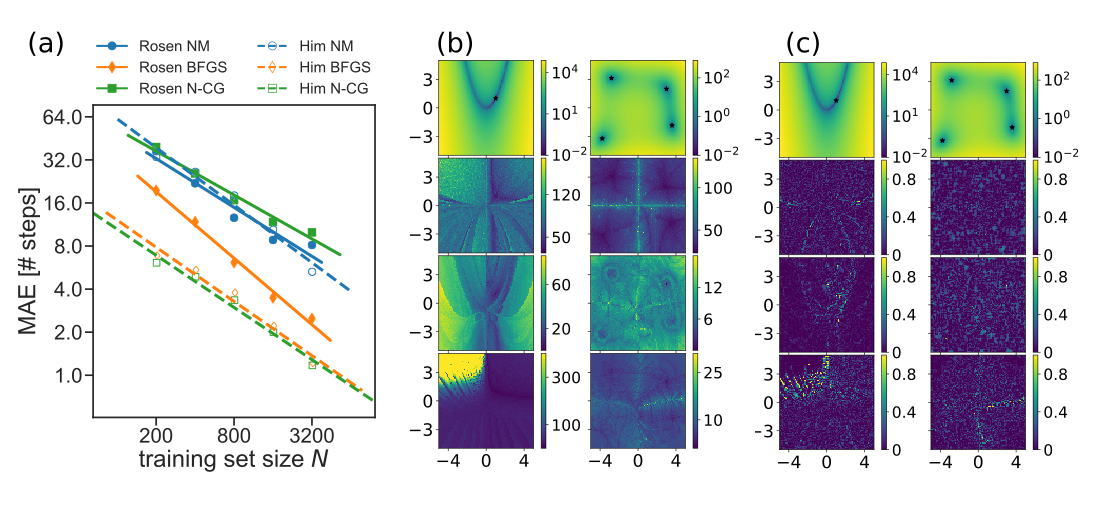}
    \caption{2D non-linear toy systems consisting of the Rosenbrock (``Rosen'') and Himmelblau (``Him'') functions and minimum search with three optimizers (Nelder-Mead (NM), BFGS, and Newton-CG (N-CG)).
    a) Learning curves showing the prediction error of KRR for Rosen (solid lines) and  Him (dashed lines) function using starting point ($x$, $y$) as representation input.
    b) Top row shows the function values for Rosen (left) and Him (right).
    Row two, three, and four show the number of optimization steps (encoded in the heat map) for 10200 starting points for NM, BFGS, and N-CG, respectively.
    c) Row two, three, and four show the relative prediction error of the ML model trained on the largest training set size $N = 3200$ for NM, BFGS, and N-CG, respectively.
    }
    \label{fig:tool_kit}
\end{figure*}
 From the total data set (10200 optimizations) 3200 were chosen randomly for every combination of optimizer and function and the prediction error was computed for different training set sizes $N$. 
Figure \ref{fig:tool_kit} a) shows the learning curves for the Rosenbrock (``Rosen'') and the Himmelblau (``Him'') functions.
Well behaved learning curves were obtained for both functions and all optimizers.
The ML models for Him-BFGS and Him-N-CG have a lower offset because the variance of the data set is smaller (between 0 and 25 optimization steps) than for the others ($\sim$50-120 steps).
The offset of Rosen-Newton-CG can be explained by the truncated runs which caused a non smooth area in the function space ($x < -0.5$ and $y > 2.5$) which leads to higher errors.
 
In addition to the learning curves, we computed the relative prediction errors of the different optimization runs.
These results are shown in Figure \ref{fig:tool_kit} c).
As expected, the errors get larger when the starting point is close to a saddle point: small changes in the starting point coordinates may lead to very different optimization paths. 
These discontinuities naturally occur for any optimizer based on the local information at the starting point and can be consistently observed in Figure \ref{fig:tool_kit} b). 
Additional discontinuities can also be observed depending on the optimizer.
For all these regions larger relative errors for KRR can be observed [shown in Figure \ref{fig:tool_kit} c)] illustrating that small prediction errors rely on a reasonably smooth target function.
In summary, we can show that KRR is capable of learning the discrete number of optimization steps which is a strong indication that the computational cost of quantum chemistry geometry optimization and transition state searches should be learnable in principle .
\begin{table*}
\scalebox{0.8}{
\begin{tabular}{l | c c c | c c c | c}
    \rowcolor{Gray}
    \textbf{Calculation}     & \multicolumn{3}{c|}{ \textbf{SP} } & \multicolumn{3}{c|}{ \textbf{GO} } & \textbf{TS}   \\
    \hline
    \textbf{Label}    &  \makecell{QM9$^{\textrm{SP}}_{\textrm{CC/DZ}}$} & \makecell{QM9$^{\textrm{SP}}_{\textrm{CC/TZ}}$} & \makecell{QMspin$^{\textrm{SP}}_{\textrm{MRCI}}$}  & \makecell{QM9$^{\textrm{GO}}_{\textrm{B3LYP}}$} & \makecell{QMrxn$^{\textrm{GO}}_{\textrm{MP2}}$} &  \makecell{QMspin$^{\textrm{GO}}_{\textrm{CASSCF}}$}  & \makecell{QMrxn$^{\textrm{TS}}_{\textrm{MP2}}$}\\
    \rowcolor{Gray}
    \textbf{$\mathbf{\textit{N}}_{\textrm{max}}$} & 5000 & 3200 & 2000 & 3200 & 6400 & 1200 & 1000 \\ 
    \textbf{BoB  [\%]} & 2.0 & 3.3 & 32.7 & 42.5 & 40.5 & 47.8 & 32.9 \\ 
    \rowcolor{Gray}
    \textbf{FCHL [\%]} & 1.3 & 1.6 & 30.9 & 37.6 & 38.9 & 39.8 & 27.0 \\ 
\end{tabular}}
\caption{QML results (normalized prediction errors) for seven task and both representations (BoB and FCHL) for largest training set size (\textbf{\textit{N}}$_{\textrm{max}}$).}
\label{tab:results}
\end{table*}
%
%We performed machine learning of wall times for the seven tasks shown in Figure \ref{fig:aTOg}. % shown in Figure \ref{fig:aTOg}).
%Figure~\ref{fig:aTOg} shows the wall time learning curves for the three different use cases (SP, GO, and TS) with the seven tasks. 
%data sets: elementary organic chemistry reactions (\textbf{A}, \textbf{B}), closed shell molecules (\textbf{C}, \textbf{D} and \textbf{E}), and strongly correlated systems (\textbf{F} and \textbf{G}).
%All measures of computational cost were normalized with respect to the median of their test set.
%
%Learning was achieved for both representations (BoB and FCHL) and in almost all tasks the more sophisticated representation FCHL outperforms BoB.
%For the FLOP count of task \textbf{QM9$^{\textrm{SP}}_{\textrm{CC/DZ}}$}, as well as
%For the tasks \textbf{QMspin$^{\textrm{SP}}_{\textrm{MRCI}}$} and \textbf{QMrxn$^{\textrm{GO}}_{\textrm{MP2}}$}, both representations perform similarly.
\subsection{Quantum Machine Learning}

\subsubsection{Single Point (SP) Wall Times}
\begin{figure*}
    \centering
    \includegraphics[width=0.9\textwidth]{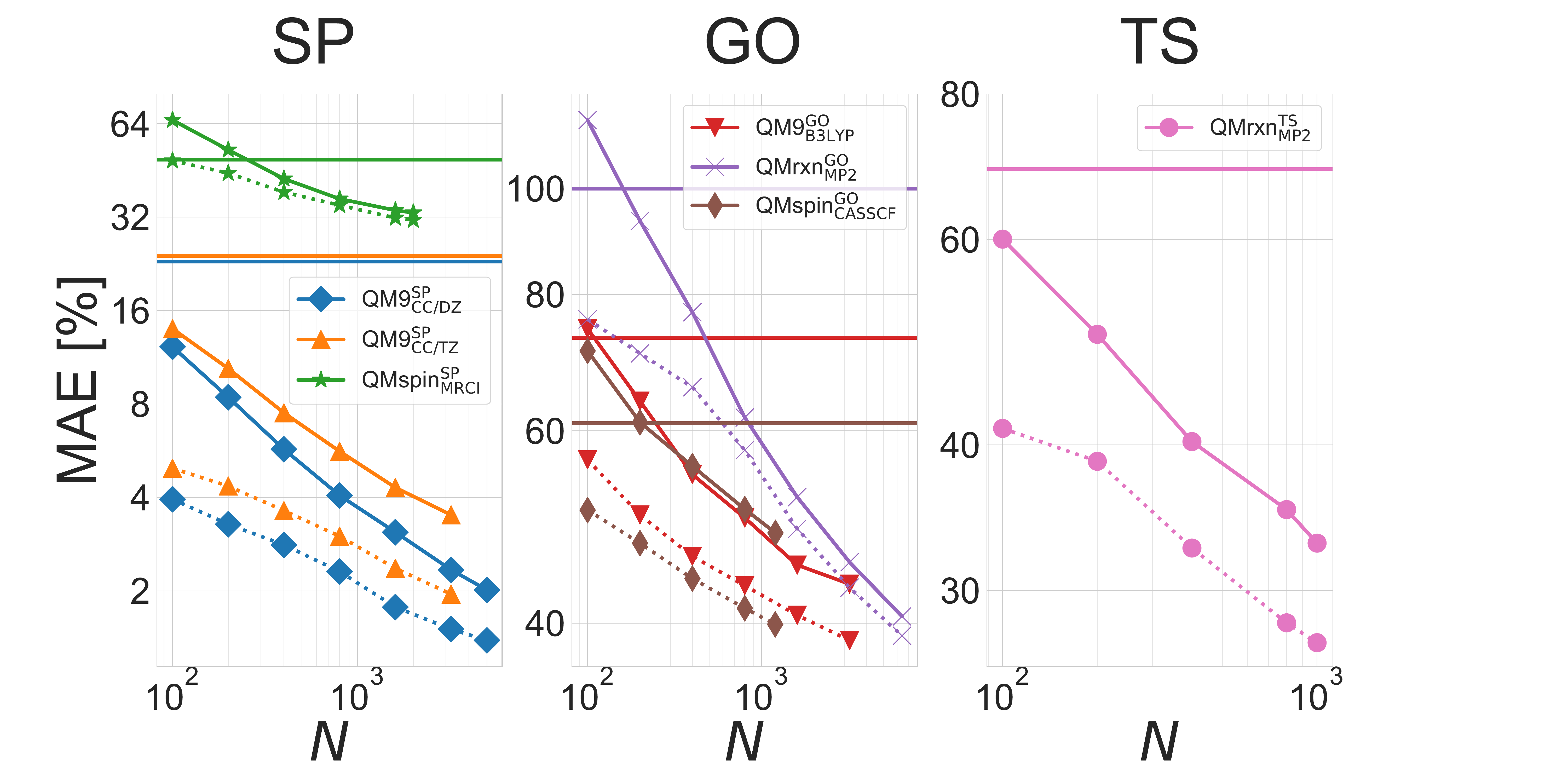}
    \caption{Learning curves showing normalized test errors (cross validated MAE divided by median of test set) for seven tasks using BoB (solid) and FCHL (dashed) representations. The model was trained on wall times normalized w.r.t. number of electrons.
Horizontal lines correspond to the performance estimating all calculations have mean run time (standard deviation divided by mean wall time of the task).}
    \label{fig:aTOg}
\end{figure*}
%

%Another reason for the worse performance of BoB is the lack of uniqueness.
In the following, learning of the wall times for the different quantum chemistry tasks is discussed, the learning of the corresponding CPU times has also been investigated and results of the latter are given in the SI.
Figure \ref{fig:aTOg} (left) shows the performance of QML models of wall times using learning curves for the SP use case. For the two similar tasks \textbf{QM9$^{\textrm{SP}}_{\textrm{CC/DZ}}$} and \textbf{QM9$^{\textrm{SP}}_{\textrm{CC/TZ}}$}, the timings of the smaller basis set was consistently easier to learn, i.e.~smaller training set required to reach similar predictive accuracy. 
Similarly to physical observables~\cite{FCHL}, the use of the FCHL representation results in systematically improved learning curve off-set with respect to BoB. 
It is substantially more difficult to learn timings of multi-reference calculations (task \textbf{QMspin$^{\textrm{SP}}_{\textrm{MRCI}}$}),
nevertheless, learning is achieved, and BoB initially also exhibits a larger off-set than FCHL, but the learning curves of the respective two representations converge for larger training set sizes. 
More specifically, for training set size $N =$ 1'600, BoB/FCHL based QML models reach an accuracy of 3.1/1.8, 4.3/2.4, and 33.7/31.8 \%
for \textbf{QM9$^{\textrm{SP}}_{\textrm{CC/DZ}}$}, \textbf{QM9$^{\textrm{SP}}_{\textrm{CC/TZ}}$}, and \textbf{QMspin$^{\textrm{SP}}_{\textrm{MRCI}}$}, respectively.
Corresponding respective average wall times in our data-sets, distributions shown in Fig.~\ref{fig:dens}, average at $\sim$6, 15, and 480 minutes. 
%The difference in the distributions (variance) of the data sets is related to the average run time, the higher the run time is the larger the variance and the more difficult learning becomes.
%On one hand, the tensor calculus in the algorithms that determine the run time are defined with respect to these non-local orbitals and the latter may not be optimally represented by a one-, \mbox{two-,} or three-body representation based on atomic centers.
%This would explain the somewhat poor performance of our QML models in predictions of these run times.
%In addition, heterogeneous compute architecture was employed for these data sets (see Table 1 in the SI). 
%On the other hand, the tensor contractions in the local coupled cluster algorithm (task \textbf{QM9$^{\textrm{SP}}_{\textrm{CC/DZ}}$} and \textbf{QM9$^{\textrm{SP}}_{\textrm{CC/TZ}}$}) are sensitively linked to the chemically relevant many-body interactions expressed in the basis of localised orbitals.
%Therefore, the computational cost may in this case be suitably encoded by the representations used.
%Furthermore, the computational cost is much less ``noisy'' and results were obtained on homogeneous hardware (cf.\ section \ref{subsec:flop}).
To the best of our knowledge, such predictive power in estimating compute timings has not yet been demonstrated for common quantum chemistry tasks.

The extraordinary accuracy that our model can reach in the prediction of the wall times for the \textbf{QM9$^{\textrm{SP}}_{\textrm{CC/DZ}}$} and \textbf{QM9$^{\textrm{SP}}_{\textrm{CC/TZ}}$} quantum chemistry tasks may be explained by the undlying quantum chemical algorithm.
The tensor contractions in the local coupled cluster algorithm are sensitively linked to the chemically relevant many-body interactions expressed in the basis of localized orbitals.
Therefore, the computational cost can be suitably encoded by atom-based machine learning representations.

\begin{figure}[!ht]
    \centering
    \includegraphics[width=.5\textwidth]{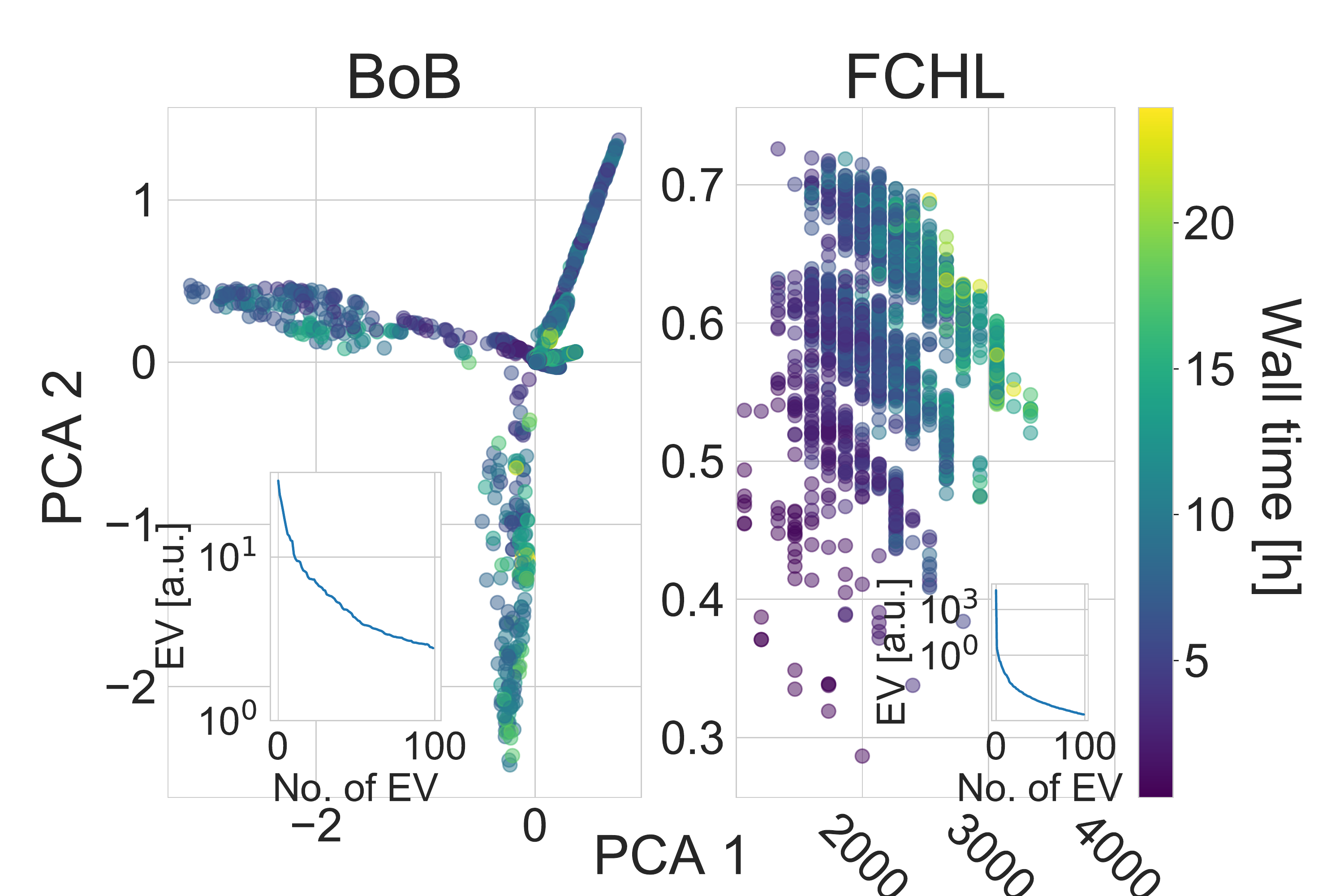}
    \caption{PCA plots of kernel elements for BoB (left) and FCHL (right) for data set \textbf{QMspin$^{\textrm{SP}}_{\textrm{MRCI}}$}.
    The weights of the two first principal components for the molecules in the data sets are plotted against each other and corresponding wall times are encoded as a heat map. Insets show the first 100 eigenvalues on a log scale.}
    \label{fig:pca}
\end{figure}

In order to investigate the relative performance of BoB vs. FCHL further, we have 
performed a principal component analysis (PCA) on the respective kernels (training set size $N =$ 2'000) for task \textbf{QMspin$^{\textrm{SP}}_{\textrm{MRCI}}$}.
The projection onto the first two components is shown in 
Figure \ref{fig:pca}, color-coded by the training instance specific wall times, and
with eigen-value spectra as insets. 
For FCHL, the decay of the eigenvalues is very rapid (tenth eigenvalue already reaches 0.1). 
From the PCA projection, the number of heavy atoms emerges as a discrete spectrum of weights for the first principal component. The second principal component groups constitutional isomers.
This reflects the importance of the one-body terms in the FCHL representation.
The data covers well both components and the color various monotonically.
All of this indicates a rather
low dimensionality in the FCHL feature space which facilitates the learning.
The kernel PCA plot of the FCHL representation shows that the learning problem is smooth in representation space and that there is a correlation between the property (computational cost) and the representation space.
By contrast, the BoB's PCA projection onto the first two components displays a star-wise pattern with
linear segments which indicate that more dimensions are required to turn the data into a monotonically varying
hypersurface.
The eigenvalue spectrum of BoB decays much more slowly with even the 100$^{th}$ eigenvalue still far above 1.0. All of this indicates that learning is more difficult, and thereby explains the comparatively higher off-set. 

%The eigenvalue spectra for the FCHL and BoB representations show a strong hierarchy of principal components and a larger number (several tens to a few hundreds) of principal components with eigenvalues of similar order of magnitude, respectively.

% \TDH: Do PCA also for data set G, which might help to explain why BoB performs here better.\\
% \TDH: Give x and y axis values for the PCA.\\
% \TDH: Maybe a plot of the weights of the BoB representation elements of the two first principal components (as obtained by the left eigenvectors of a SVD on the BoB representation matrix) is more useful than a PCA on the BoB kernel. \\
% \TDH: Maybe give one formula on how the PCA is computed?\\
% \TDH: The PCA analysis is needed with two more heat maps: total number of atoms in the molecule and clustering of the constitutional isomers.\\
% \TDH: We should also try SLATM for the MRCI case in order to understand where the bad performance of FCHL comes from. \\
%
\subsubsection{Geometry Optimization (GO) Wall Times}
Learning curves in Figure \ref{fig:aTOg} (middle) shows that it is, in general, possible to build QML models of GO timings for the tasks considered. 
We obtained accuracies for BoB/FCHL for $N$ = 800 of 50.0/43.3, 61.7/57.6, and 50.7/41.2\% for tasks \textbf{QM9$^{\textrm{GO}}_{\textrm{B3LYP}}$}, \textbf{QMrxn$^{\textrm{GO}}_{\textrm{MP2}}$}, and \textbf{QMspin$^{\textrm{GO}}_{\textrm{CASSCF}}$}, respectively.

\begin{figure}[ht!]
    \centering
    \includegraphics[width=0.5\textwidth]{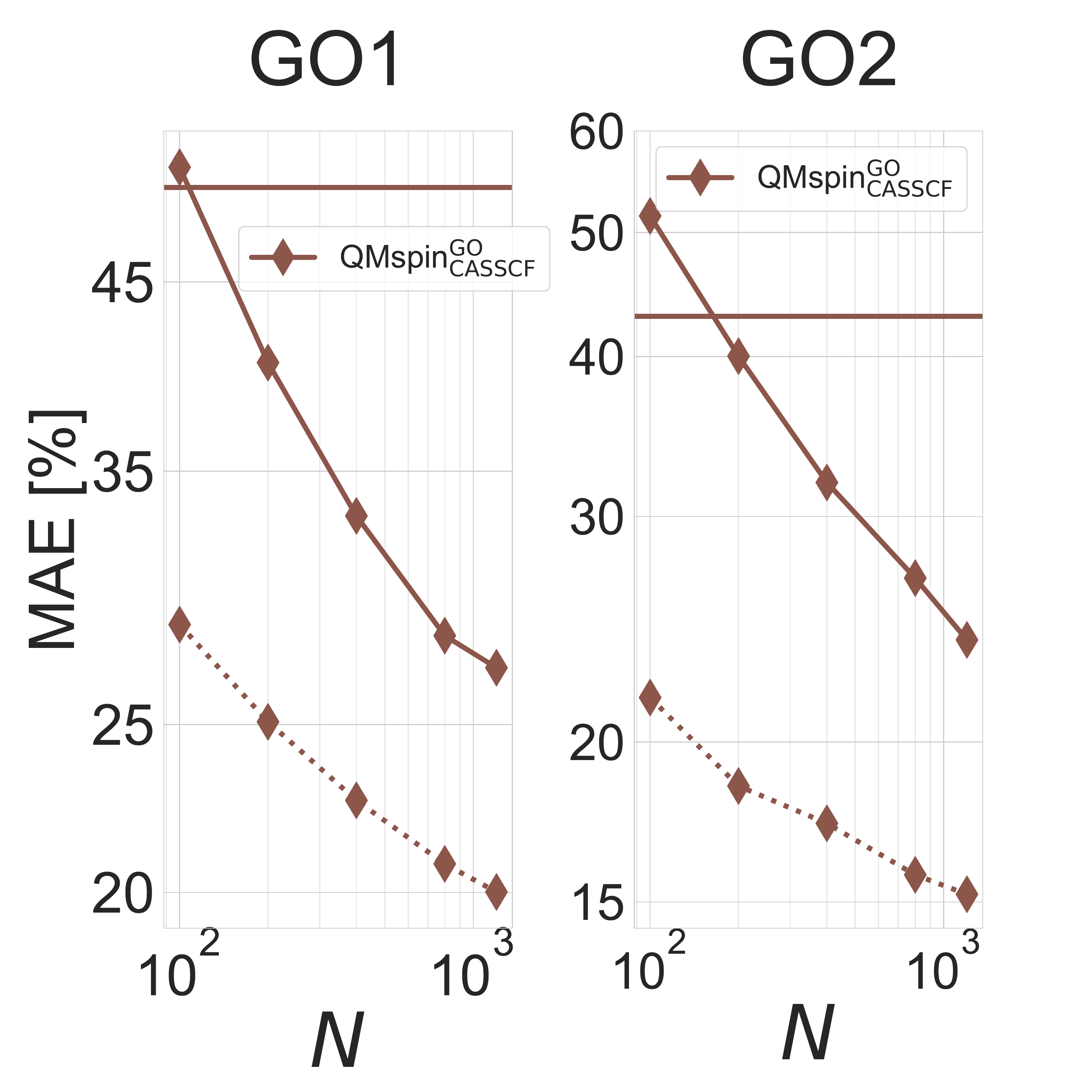}
    \caption{Learning curves showing normalized test errors (cross validated MAE divided by median of test set) for the first two geometry optimization steps on task \textbf{QMspin$^{\textrm{GO}}_{\textrm{CASSCF}}$} using BoB and FCHL as representations. The model was trained on CPU times divided by the number of electrons. Horizontal lines correspond to the performance estimating all calculations have mean run time (standard deviation divided by the mean wall time of the data set).}
    \label{fig:go}
\end{figure}

Interestingly, the comparatively larger off-set in the learning curves, however, indicates that it is more difficult to learn GO timings than SP timings. This is to be expected since GO timings involve not only SP calculations for various geometries but also geometry optimization steps.
%Although the learning of GO is a much harder problem to learn it also depends on the method used.
%For example it was more difficult to learn SP timings for MR (F) and CASSCF (G) calculations because of I/O over head. %The closer the initial geometry is to the target stationary point geometry, the faster it will converge and the less steps are required which results in a shorter run time.
In other words, the QML model has to learn the quality of the initial guesses for subsequent GO optimizations. This can not be expected to be a smooth function in chemical space. Furthermore, 
the mapping from an initial geometry (used in the representation for the QML model) 
to the target geometry can vary dramatically when the initial geometry happens to be close to a saddle point (or a second order saddle point in the case of TS searches, see next section): Very slight changes in the initial geometry (or in the setup of the geometry optimization) may lead to convergence to very different stationary points on the potential energy surface.
This makes the statistical learning problem much less well conditioned than for single point calculations, which also reflects in the larger variance of the geometry optimization timings compared to single point calculations.
As such, GO timings represent a substantially more complex target function to learn than SP timings. 
Note that for  any task (even for the toy system applications) we require a different QML model.
The cost of the GO depends on the initial geometry and the convergence criteria.
The latter varies only slightly within a data set.
The former is part of the representation of the molecular structure and therefore captured by our model. 
The input structures for the task \textbf{QMrxn$^{\textrm{GO}}_{\textrm{MP2}}$} are derived from the same molecular skeleton and are therefore very similar.
The same holds for task \textbf{QM9$^{\textrm{GO}}_{\textrm{B3LYP}}$} and \textbf{QMspin$^{\textrm{GO}}_{\textrm{CASSCF}}$} which are derived from QM9 molecules.
The convergence criteria also stay the same for all calculations within a data set and would only cause a more difficult learning task if a machine was trained over several different data sets.
We also showed with the toy system that it is possible to learn the number of steps for different optimizer starting from different areas on the surface (see Figure 4 b)).
%(cf.\ horizontal lines in Figure~\ref{fig:aTOg}).
%%%%%%%%%%%%%%%%%%%%%%%%%%%%%%%%%%%%%%%%%%%%%%%%%%%%%%%%%%
% maybe it should go into section quantum machine learning
%%%%%%%%%%%%%%%%%%%%%%%%%%%%%%%%%%%%%%%%%%%%%%%%%%%%%%%%%%
%Similar challenges were addressed by Kulik and coworkers\cite{KulikFailPred} when predicting the successful outcome of DFT geometry optimizations of transition metal species as a classification problem.
%When using support vector machines and ANN with connectivity-only representations as input they were able to optimize their job list by rather reliably discarding jobs that will not properly converge for the given method.
%They could further improve their results by using representations containing electronic and geometric features obtained during the geometry optimization.
%to improve the predictions of geometry optimizations for the task \textbf{GO$^{\textrm{QMspin}}_{\textrm{CASSCF}}$}, we splitted the individual optimization steps into the first step (\textbf{GO1}) and the subsequent steps (\textbf{GO2}), because the first step takes in average $\sim$20\% more time than the following steps (for more details we refer to section 1.4 of the SI).
%which are roughly on the same order of magnitude.
%The reason for this behaviour is that Molpro uses, for a CASSCF geometry optimization, the wave function of a previous step as an initial guess for the CASSCF wave function of the new geometry.
%For this reason, the first step takes significantly longer then the following steps.
To further improve the performance of our model of task \textbf{QMspin$^{\textrm{GO}}_{\textrm{CASSCF}}$}, we split the GO into the first GO step (GO1) and all subsequent steps (GO2). 
This choice has been motivated by our observation that most of the variance stemmed from the first
GO step (requiring to build the wave-function from scratch), while the subsequent steps for themselves
have a substantially smaller variance. 
The resulting learning curves are shown in Figure~\ref{fig:go} and justify this separation in leading to an improvement of the QML model to reach errors of less than 25\% at $N$ = 800 (rather than more than 40\%), as well as further improved job scheduling optimization (shown below in Figure \ref{fig:gostepscheduling}). 

%Local correlation methods are based on the locality of the molecular orbitals.
%In this case, the localized occupied orbitals may closely resemble the molecular orbitals that chemists employ for conceptual chemical reasoning and virtual orbitals are also of limited spacial extend.
%This is however not the case in non-local methods, such as the CASSCF and MRCI calculations presented here.
%In the latter, the spatial extend of the molecular orbitals strongly depends on orthogonalization tails that arise from the non-orthogonality of the atomic basis set.
\subsubsection{Transition State (TS) Wall Times}
Transition state search timings were slightly easier to learn than geometry optimization timings (see Figure \ref{fig:aTOg} (right)).
Particularly for the larges training set size (\textbf{$N_{max} = 1000$})  for BoB/FCHL we obtained MAEs of 32.9/27.0\% and reduced the off-set by $\sim 10\%$ compared to learning curves for the GO use case. 
As already discussed in the previous section, the run time of GO and TS timings not only scales with the number of electrons but also depends on the initial structure.
For the transition state search, the scaffold (which is close to a transition state) was functionalized with the different functional groups.
Since the initial structures were closer to the final TS the offset of the learning curves is lower than for learning curves of the GO use case, where the initial geometries were generated with a semi empirical method (PM6) for task \textbf{QMrxn$^{\textrm{GO}}_{\textrm{MP2}}$}, carbenes were derived from QM9 molecules for task \textbf{QMspin$^{\textrm{GO}}_{\textrm{CASSCF}}$}, and geometries for task \textbf{QM9$^{\textrm{GO}}_{\textrm{B3LYP}}$} were obtained with a different basis set.

A summary of the results for all tasks for the largest training set size (\textbf{\textit{N}}$_{\textrm{max}}$) can be found in Table \ref{tab:results}.
\subsubsection{Timings, Code, Hardware}
 Regarding hardware dependent models, within one data set we only used one electronic  structure code which is also consistent with the general handling of the data set generation.
    The noise that is generated using different infrastructures affects the learning only in a negligible amount  in our case, since the difference in hardware capabilities is minimal.
    When looking at the task \textbf{QMrxn$^{\textrm{TS}}_{\textrm{MP2}}$} where we used five different CPU types on two clusters (Table 1 in the SI), we could not find any evidence that different hardware affects the learning compared to other GO tasks that ran on only one CPU type and cluster.
    However the hardware for these calculations is still very similar.
    When it differs to a greater extant, the noise level will rise.
    The noise does not only depend on the cluster itself but also on other calculations running on the cluster which is non-deterministic and will limit the transferability of the ML models.
    For this reason we removed some of the timings with large I/O overhead using Figure 3.
    For the \textbf{QM9$^{\textrm{SP}}_{\textrm{CC}}$} tasks, the run time difference using the Intel MKL 2019 library\cite{MKL} and OpenBlas 0.2.20\cite{OpenBlas} were computed for a few cases and are found to be only within a few percents of the wall time.
    Furthermore, run times of a native build of the Molpro software package version 2018.3 with OpenMPI 3.0.1\cite{OpenMPI}, GCC 7.2.0\cite{GCC}, and GlobalArrays 5.7\cite{GAWeb,GAPub} and the shipped executable were compared and yielded run times within a few percents of difference.
    The FLOP calculations on the \textbf{QM9$^{\textrm{SP}}_{\textrm{CC}}$} data set have been performed on a compute node with 24 processors [Intel(R) Xeon(R) CPU E5-2650 v4 @ 2.20GHz (Broadwell)].
    The significant part of the FLOP clock cycles constituted of vectorized double precision FLOP on the full 256 bit FLOP register, i.\ e.\ the essential numerical operations of the quantum chemistry algorithm were directly measured. Hence, FLOP count constitutes a valuable measure of the compute cost in our case.\cite{FLOPSrate2}
    We anticipate that Hardware specific QML models will be used in practice.
\subsubsection{Single Point (SP) FLOPs}
\begin{figure}[ht!]
    \centering
    \includegraphics[width=0.45\textwidth]{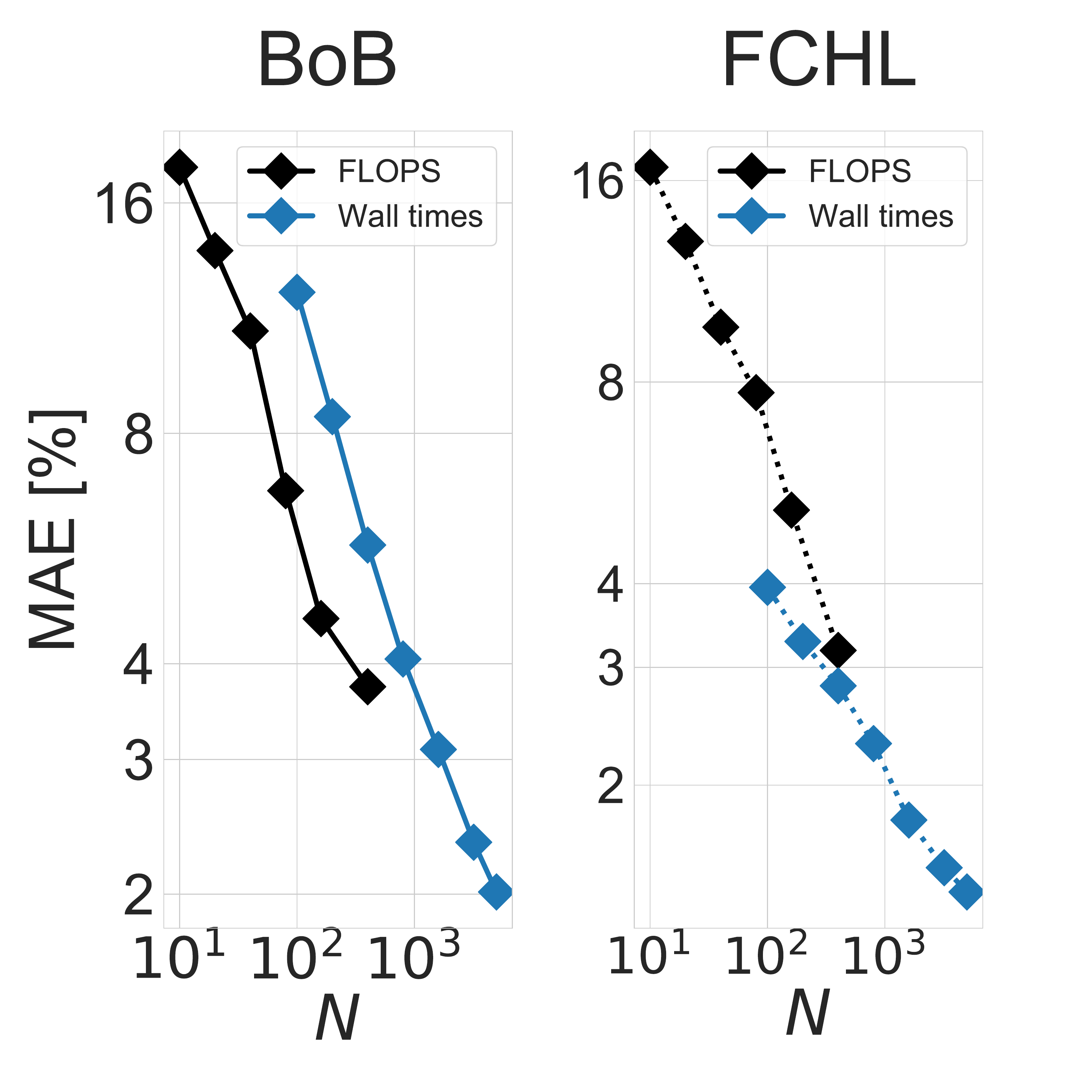}
    \caption{Learning curves showing normalized prediction errors (cross validated MAE divided by median of test set) for FLOP count and wall times on task \textbf{QM9$^{\textrm{SP}}_{\textrm{CC/DZ}}$} using BoB and FCHL representations.}
    \label{fig:flops}
\end{figure}

% The first step in testing our models was to generate learning curves to evaluate the performance of our models for the different types of calculations (SP, GO, TS and GO steps).\\\\
To provide unequivocal numerical proof that it is justifiable to learn wall times we applied our models to FLOP counts for the task \textbf{QM9$^{\textrm{SP}}_{\textrm{CC/DZ}}$}, shown in Figure \ref{fig:flops}.
FLOP count as a ``clean'' measurement (almost no noise) for computational cost was slightly easier to learn than wall times and the learning curves show similar behaviour:
The model trained on the same task \textbf{QM9$^{\textrm{SP}}_{\textrm{CC/DZ}}$} reaches $\sim$4\% MAE already with just 400 training samples, while $\sim$1000 training samples were required in the case of wall times using BoB. For FCHL, the performance is similar but the slope is steeper for the FLOP model which indicates a faster learning or less noise.

\subsection{Application: Optimal Scheduling}
\subsubsection{Job Array and Job Steps}
\begin{figure*}
    \centering
    %,trim=1.5cm 0.5cm 2cm 0cm, clip
    \includegraphics[width=\textwidth]{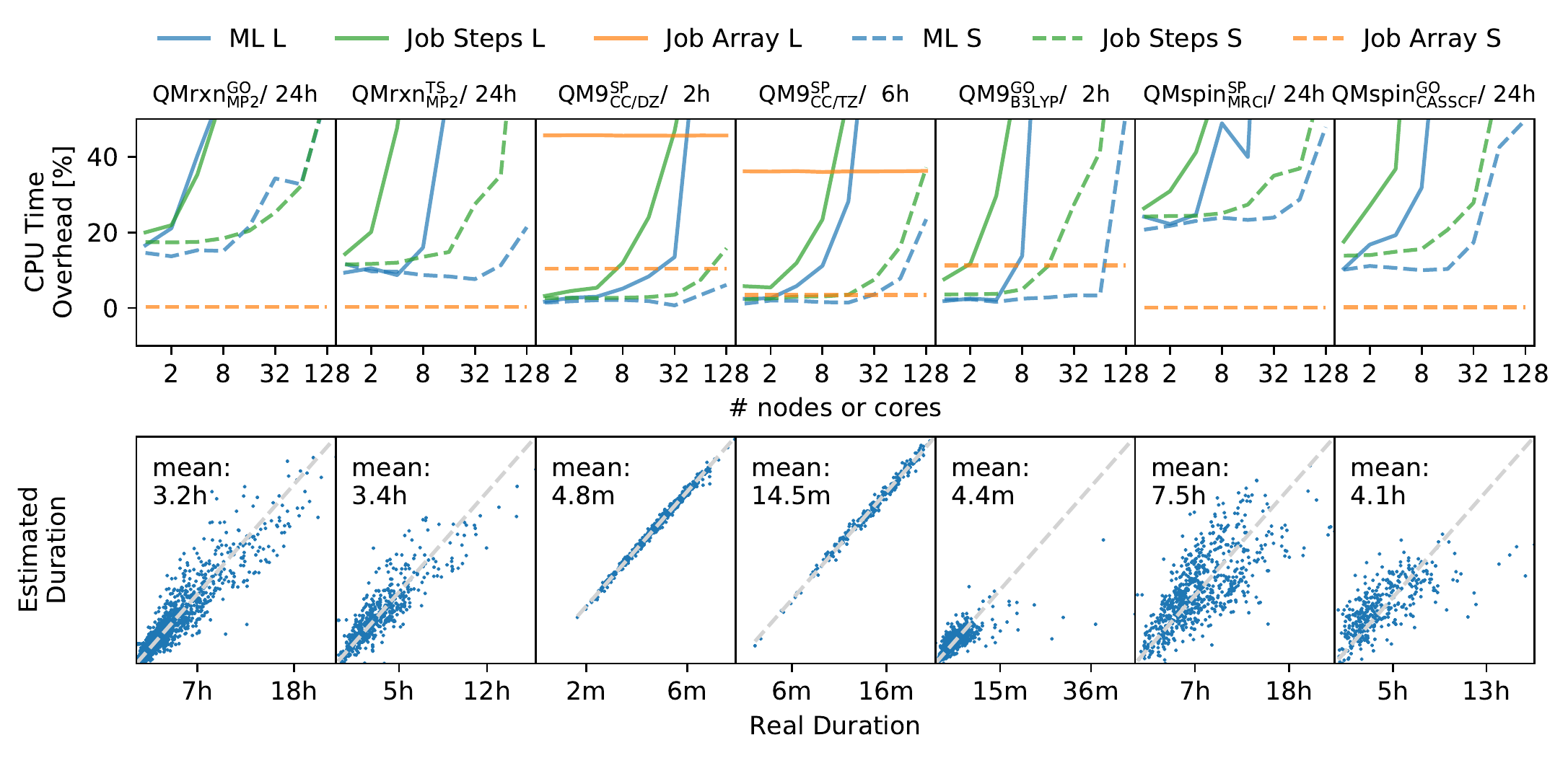}
    \caption{Scheduling efficiencies for the seven different tasks (columns) assuming a certain per-job wall time limit specified in column title.
Infrastructure assumptions correspond to either a large (solid lines, L) compute center or a small (dashed lines, S) university compute center.
% PLEASE LEAVE THIS COMMENT HERE
%Wall time (first row) for established methods (job array and jobs steps, see text) normalized to wall time of our suggested method (ML) to show the comparison on an homogeneous scale.
%Wall time ratios of larger than one mean our method is more efficient.
Top row reports CPU time overhead reduction when using the QML based (blue) rather than the conventional (green, orange) packing. Results are given relative to the total CPU time needed for the calculations of each data set for established methods (job array and jobs steps, see text) and our suggested method (QML).
Bottom row shows actual vs. predicted times (using FCHL as representation) for all calculations in each data set using maximum training set size.}
    \label{fig:scheduling}
\end{figure*}

%\subsection{Job Scheduling}
For the scheduling optimization for all seven tasks (\textbf{QM9$^{\textrm{SP}}_{\textrm{CC/DZ}}$}, \textbf{QM9$^{\textrm{SP}}_{\textrm{CC/DT}}$}, \textbf{QMspin$^{\textrm{SP}}_{\textrm{MRCI}}$}, \textbf{QM9$^{\textrm{GO}}_{\textrm{B3LYP}}$}, \textbf{QMrxn$^{\textrm{GO}}_{\textrm{MP2}}$}, \textbf{QMspin$^{\textrm{GO}}_{\textrm{CASSCF}}$}, \textbf{QMrxn$^{\textrm{TS}}_{\textrm{MP2}}$}), the QML model with the best representation (lowest MAE with maximum number of training points) was used which in all cases was FCHL.
   For the FFD algorithm absolute timing predictions are needed to make good decisions.
    The lower panel of Figure 9 shows the accuracy of the QML predictions.
    While the individual predictions (absolute not relative) are in many cases not perfect and partially still exhibit a significant MAE (cf. Figure 5), this level of accuracy is already sufficient to reduce the overhead of the job scheduling.
The lower panel of Figure~\ref{fig:scheduling} shows the accuracy of the QML predictions.
While the individual predictions (absolute not relative) are in many cases not perfect and partially still exhibit a significant MAE (cf.\ Figure~\ref{fig:aTOg}), this level of accuracy is already sufficient to reduce the overhead or the wall time limits of the job scheduling.
In particular, in the limit of a large number of cores working in parallel, our approach typically halved the computational overhead (data sets with closed shell systems and TS searches) while also reducing the time to solution by reducing the total wall time.
This shows that for the scheduling efficiency problem, it is not required to obtain perfect estimates for the individual job durations, but rather reasonably accurate estimates.
However, if there was the need for better accuracy, by virtue of the ML paradigm (prediction error decay systematically with training set size) this could easily be accomplished by decreasing the error simply through the addition of more training data. 
%This is particularly true since any bin packing algorithm employed in a later stage will be of heuristic nature since the run time complexity for the optimal solution is too high.

When comparing the different methods in the upper panel of Figure~\ref{fig:scheduling}, we see that the job array approach had no overhead for cases where single-core jobs can be submitted separately.
While this is true it means that every job needs to wait in the queue again, thus increasing the total time to solution.
For large task durations, this effect is less pronounced but typically the job array approach doubles the wall time which renders this approach unfavourable. 
%In addition, large compute clusters normally do not offer the possibility of submitting single core jobs.

Using job steps alone becomes inefficient if the task durations are long, since the assumption that all tasks are roughly of identical duration will mean that interruptions of unfinished calculations occur more often.
Having a more precise estimate allows for more efficient packing.
This becomes important on large compute clusters where only full nodes can be allocated: In this case, the imbalance of the durations of calculations running in parallel further increases the overhead.
Our method typically gave a parallelization overhead of 10-15\%\ for a range of data sets.
For example, in the task \textbf{QMrxn$^{\textrm{GO}}_{\textrm{MP2}}$}, our approach allowed us to go to two orders of magnitude more compute resources and have the same overhead as job step parallelization.
%which results in a saving of  33M US\$/year (shown in Figure \ref{fig:cscs_usage}).
This is a strong case for using QML based timing estimates in a production environment -- in particular, since the number of training data points required is very limited (see Figure~\ref{fig:aTOg}).
\begin{figure}
    \centering
    \includegraphics[width=\columnwidth]{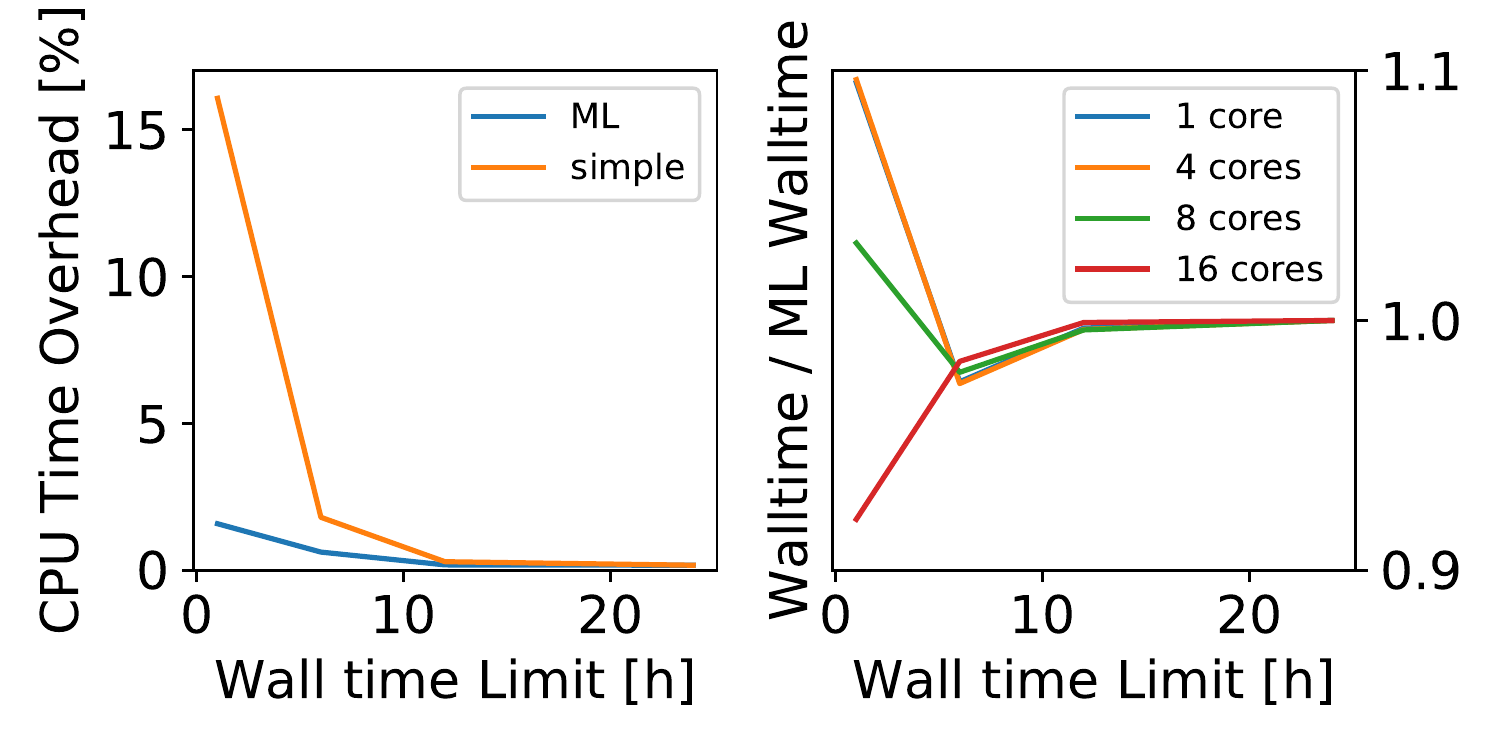}
    \caption{CPU time overhead and wall time for geometry optimizations compared between the simple approach and the QML approach.
See text for details of the strategies.
CPU time overhead given in percent relative to the bare minimum of CPU time needed.
Wall time given relative to the wall time resulting from using the QML approach.
  All geometry optimizations come from task \textbf{QM9$^{\textrm{GO}}_{\textrm{CASSCF}}$}.}
\label{fig:gostepscheduling}
\end{figure}

\subsubsection{Geometry Optimization Steps}
Given that the number of steps of a geometry optimization is difficult to learn (see lower panel of Figure~\ref{fig:scheduling}), the ability to accurately predict the duration of a single geometry optimization step allows to increase efficiency via another route.
On hybrid compute clusters, the maximum duration of a single compute job is limited.
We suggest to check during the course of a geometry optimization whether the remaining time of the current compute job is sufficient to complete another step.
If not, it is more efficient to relinquish the compute resources immediately rather than committing them to the presumably futile undertaking of computing the next step.
We refer to these strategies as the ``simple approach'' (take all CPU time you can, give nothing back) and the ``QML approach'' (give up resources early).
Figure~\ref{fig:gostepscheduling} shows the advantage of the QML approach: it allows to go towards shorter compute jobs and reduces the CPU time overhead by up to 90\% for small wall time limits using the job array approach.
This is more efficient for the scheduler and increases the likelihood of the job being selected by the backfiller, further shortening the wall time.
%We can reduce the computational overhead in this example to less than one third of the traditional approach.
Using the QML approach does not severely affect the wall time, i.e. the time-to-solution.
This is largely independent of the extent of parallelization employed in the calculation (see right hand side plot in Figure~\ref{fig:gostepscheduling}).
We suggest to implement an optional stop criterion in quantum chemical codes where an external command can prematurely stop the progress of the geometry optimization to be resumed in the next compute job.
This change can drastically improve computational efficiency on large scale projects.
Estimating the current consumption to be on the order of at least 5$\cdot 10^5$ petaFLOPS (see discussion above in section \ref{sec:intro}) for computational chemistry and materials science this 
approach may lead to potentially large savings in economical cost.
%
%%%%%%%%%%%%%%%%%%%%%%%%%%%%%%%%%%%%%%%%%%%%%%%%%%%%%%%%%%%
% Conclusion
%%%%%%%%%%%%%%%%%%%%%%%%%%%%%%%%%%%%%%%%%%%%%%%%%%%%%%%%%%%
\section{Conclusion}
\label{sec:conclusion}

%Floating point operations (FLOP) count may be considered as the most direct measure of the computational complexity and we indeed succeed in very efficient machine learning of the FLOP count.
%In comparison, CPU times depend much more on the specific hardware and software configuration (and possibly also on external influences).
%Finally, wall times are much less deterministic, depending in some use cases strongly on external factors (e.\ g., concurrent work loads on the compute clusters), that are usually not reflected in quantum machine learning (QML) representations.
%Nevertheless, we could also obtain efficient machine learning for both CPU and wall times.

We have shown that the computational complexity of quantum chemistry calculations can be predicted across chemical space by QML models.
First we looked at a 2D non-linear toy system consisting of example functions which are known to be difficult to optimize.
Using these test functions and three optimizers, we build a first ML model and the learning curves show that it is possible to learn the number of optimization steps using only the starting position ($x$, $y$). 
Representations are designed to efficiently cover all relevant dimension in the given chemical space.
Hence, if the computational cost is learnable by QML models, it is a reasonably smooth function in the variety of chemical spaces that we considered.
This is a fundamental result.

Our approach succeeds in estimating realistic timings of a broad variety of representative calculations commonly used in quantum chemistry work-flows: single-point calculations, geometry optimizations, and transition state searches with very different levels of theory and basis sets.
The machine learning performance depends on the quantum chemistry method and on the type of computational cost that is learned (FLOP, CPU, wall time).
While the accuracy of the prediction is shown to be strongly dependent on the computational method, we could typically predict the total run time with an accuracy between 2\% and 30\%. 
%In this context, it is analyzed how the underlying physical approximations and algorithm implementation of the quantum chemistry method reflect in the ease of learning the computational cost with QML.

Exploiting QML out-of-sample predictions, we have demonstrably used compute clusters more efficiently by reordering jobs rather than blindly assuming all calculations of one kind to fit into the same time window.
Without significant changes in the time-to-solution, we reduced the CPU time overhead by 10\% to 90\% depending on the task. 
%In near the future, further large scale quantum chemical data productions will gain importance, since machine learning models can always profit from a certain amount of reliable and accurate training data.
With the scheme presented in this work, compute resource usage can be significantly optimized for large scale chemical space compute campaigns.
To support this case, all relevant code, data, and a simple-to-use interface is made available to the community online.\cite{github}.

We believe that our findings are important since it is not obvious that established QML  models, designed for estimating physical observables, are also applicable to more implicit quantities such as computational cost.
\\
\\
\begin{acknowledgement}\\
M.\ S.\ would like to acknowledge Dr.\ Peter Zaspel for helpful discussions on the FLOP measurement and the setup of the native program builds.
This work was supported by a grant from the Swiss National Supercomputing Centre (CSCS) under project ID s848. 
Some calculations were performed at sciCORE (http://scicore.unibas.ch/) scientific computing core facility at University of Basel.
We acknowledge funding from the Swiss National Science foundation (No.~407540\_167186 NFP 75 Big Data,  200021\_175747)
and from the European Research Council (ERC-CoG grant QML).
This work was partly supported by the NCCR MARVEL, funded by the Swiss National Science Foundation.
\end{acknowledgement}
\noindent\textbf{\\Data availability statement}\\
Any data (except the carbene data set) that support the findings of this study are included within the article. The carbene data set is available from the corresponding author upon reasonable request.
\bibliography{literatur}
\newpage
\end{document}

This material is available free of charge via the Internet at \texttt{http://pubs.acs.org/}. Code and raw data is available on GitHub \texttt{https://github.com/ferchault/mlscheduling}
\section{Additional details on the data sets}
Table \ref{tab:datasets_cpu} shows additional information regarding the used hardware.
\begin{table*}
\scalebox{0.8}{
\begin{tabular}{l|ccc|ccc|c}
    \textbf{Calculation}     & \multicolumn{3}{c|}{ \textbf{SP} } & \multicolumn{3}{c|}{ \textbf{GO} } & \textbf{TS}   \\
    \hline
    \hline
    \textbf{Data set}  & \multicolumn{2}{c}{ QM9 } & QMspin & QM9 & QMrxn & QMspin  & QMrxn \\
    \hline
%
%    \textbf{Software}   &  \makecell{Molpro \\2018.3}    &\makecell{Molpro \\2018.3} &  \makecell{Molpro \\2015.1} & \makecell{Molpro \\2018.3} & \makecell{ORCA \\4.0.1} &  \makecell{Molpro \\2015.1}& \makecell{ORCA \\4.0.1}\\
%    \hline
    %
    \textbf{\# Cores} &  24       &   24       &   1      &   1      &   1       &   1       &   1 \\
    \hline
    \textbf{CPU Types} &  E5-2680v3 & E5-2680v3 & \makecell{E5-2650v2  \\ E5-2640v3  \\ E5-2630v4 } & E5-2630v4 & \makecell{E5-2640v3  \\ E5-2650v4} & \makecell{E5-2650v2\\ E5-2640v3  \\ E5-2630v4 } & \makecell{E5-2650v2\\ E5-2680v3 \\ E5-2640v3\\ E5-2630v4 \\ E5-2650v4 } \\
    \hline
    $\sigma_\text{BoB}$      & 204.8   & 204.8  &    51.2   &   51.2  &   102.4  &   51.2   &   204.8 \\
    $\sigma_\text{FCHL}$     &  12.8    &  12.8    &   51.2   &   409.6   &   51.2   &   51.2   & 51.2 \\
    \hline
    $\lambda_\text{BoB}$      & 1e-7     & 1e-7    &    1e-7   &   1e-7   &   1e-7   &   1e-5    &  1e-7\\
    $\lambda_\text{FCHL}$     &  1e-7      &  1e-7     &   1e-7    &   1e-7    &   1e-9   &   1e-5    & 1e-7 \\
    \hline \hline
\end{tabular}}
\caption{Data sets of calculations used in this work: Software used for calculations, number of cores used per calculation, and CPU types the calculation ran on as well as details of the ML hyperparameter.}
\label{tab:datasets_cpu}
\end{table*}
%
%
%\subsection{MP2 geometry optimization (QMrxn$^{\textrm{GO}}_{\textrm{MP2}}$)}
\subsection{QMrxn$^{\textrm{GO}}_{\textrm{MP2}}$}
\label{subsub:mp2_opt}
The initial reactant geometries from the reaction data set were obtained by generating the unsubstituted molecule (hydrogen atoms instead of functional groups and Fluor as leaving group) without the nucleophile. Subsequently substituting the hydrogen atoms with functional groups span the chemical space.
For every reactant a conformer search on PM6-D3 level was performed using ORCA.
The lowest lying conformer geometries were then further optimized on MP2/6-31G* level of theory which resulted in the data set set \textbf{QMrxn$^{\textrm{GO}}_{\textrm{MP2}}$}.
%
%\subsection{MP2 transition state search (QMrxn$^{\textrm{TS}}_{\textrm{MP2}}$)}
\subsection{QMrxn$^{\textrm{TS}}_{\textrm{MP2}}$}
The starting geometries for the transition state (TS) search were obtained in a similar way as described in section \ref{subsub:mp2_opt}. 
A transition state search was performed on the unsubstituted case and from the found TS the chemical space was spanned by exchanging the hydrogen atoms with functional groups. The following timings (using ORCA 4.0.1) and the initial geometries of the TS search form the data set \textbf{QMrxn$^{\textrm{TS}}_{\textrm{MP2}}$}. 
%
%\subsection{DFT geometry optimization (QM9$^{\textrm{GO}}_{\textrm{B3LYP}}$)}
\subsection{QM9$^{\textrm{GO}}_{\textrm{B3LYP}}$}
The QM9 data set contains geometries optimized with B3LYP/6-31G*.
5001 out of these 134k molecules were further optimized with a larger basis set (def2-TZVP) using  Molpro to obtain data set \textbf{QM9$^{\textrm{GO}}_{\textrm{B3LYP}}$}.
%
%\subsection{Multi-reference calculations (QMspin$^{\textrm{SP}}_{\textrm{MRCI}}$ and QMspin$^{\textrm{GO}}_{\textrm{CASSCF}}$)}
\subsection{QMspin$^{\textrm{SP}}_{\textrm{MRCI}}$ and QMspin$^{\textrm{GO}}_{\textrm{CASSCF}}$}
\label{sec:mr}
For the geometry optimization of data set \textbf{QMspin$^{\textrm{GO}}_{\textrm{CASSCF}}$} we use the CASSCF single point energy\cite{Werner1985JCP} and energy gradient implementation\cite{Busch1991JCP} in Molpro.
The calculations have been run on one compute core per job and similar amounts of run time are spent for the wave function computation and the energy gradient.
When performing a geometry optimization, the CASSCF wave function of a previous step is used as a starting guess for the CASSCF wave functions of the new geometry.
For that reason, the first step of a geometry optimization takes significantly longer than the following steps.
We take this aspect into account in our ML model as well as in our scheduling model.

%The data set F has been obtained with the internally contracted MRCISD(Q)-F12\cite{Knowles1988CPL,Werner1988JCP,Shiozaki2011JCP} method in Molpro.
%The significant part of the MRCI run time was spent in the computation of the single-pair interactions, the pair-pair interactions, as well as the F12 contributions.
%All MRCISD schemes that are based on the Davidson algorithm\cite{Davidson1975JCoP} scale formally as $n^2\cdot m^4$, where $n$ the number of correlated occupied orbitals and $m$ the number of basis functions.\cite{SherrillScalCI}
%We also consider this scaling behaviour in the setup of our ML model.
%
%All timings of multireference calculations were obtained with the shipped Molpro 2015.1 version, the calculations have been performed on somewhat heterogeneous sciCORE hardware as outlined in Table~\ref{tab:datasets}.
%I/O operations on the local hard disk drive took a non-negligible part of the compute time, an aspect that has also been studied in our ML models. \\
%
%
\section{First fit decreasing algorithm}
The bin packing problem is NP-hard\cite{Garey1990}, i.\ e., the search for the optimal solution to the problem is prohibitively expensive for real-world workloads of thousands of jobs even if the time estimates were arbitrarily accurate\cite{Martello1990,Martello1990b,Korf2002}.
First Fit Decreasing (FFD) is one of the many heuristic algorithms\cite{Coffman1984} that exists for the bin packing problem.
It has been shown that for practical purposes the FFD algorithm is close to the optimal solution, as for $q$ compute jobs as it uses at most $11/9q+1$ jobs\cite{Yue1991}, but typically is within a few percent of the optimal solution\cite{Korf2002}.
In all cases, we calculate the total core hours and the total duration from the first to the last job.
The total core hours divided by the sum of the real run times define the compute overhead.
%Ideally, the two numbers are identical, but in practice parallelization inefficiencies give rise to a significant overhead.
The total duration from first to last job should not be exceedingly high compared to other approaches, since this metric is about enabling science: if the calculations would take too long, a research project would not be started.
The ideal approach therefore reduces the overhead while keeping the total wall time at least comparable to established approaches. 

%Some of the calculations we considered are interruptible, e.\ g., geometry optimizations which can be restarted from any intermediate molecular geometry.
%Using machine-learned predictions of the individual steps of a set of geometry optimizations, we can decide on the fly whether the remaining time of a given allocation on a compute cluster is sufficient to complete the next step of the calculation.
%In the traditional approach, a given wall time limit is chosen for all calculations of one type, and the few ones that do not fit into this wall time limit are resubmitted with a higher wall time limit.
%%This is to allow for shorter jobs to be submitted, which is more efficient from the scheduler perspective.

\newpage
\section{Learning curves}
In the following we compare models with respect to different training inputs. We trained models on CPU, normalized (by number of electrons) CPU, wall, and normalized wall times. For CPU times only calculations done with Molpro could be considered because ORCA output files only contain total (wall) times. Best performance was reached with models trained on normalized CPU times. The differences are small (around 1\% to 4\%). For predictions normalized wall times were used because of their application to the scheduling.
For the test sets \textbf{QMspin$^{\textrm{SP}}_{\textrm{MRCI}}$} and \textbf{QMspin$^{\textrm{GO}}_{\textrm{CASSCF}}$} we also training on CPU times normalized by the formal scaling of the method, this did neither lead to significant changes in the training model (results not shown).
\begin{figure}[!ht]
    \centering
    \includegraphics[width=.5\textwidth]{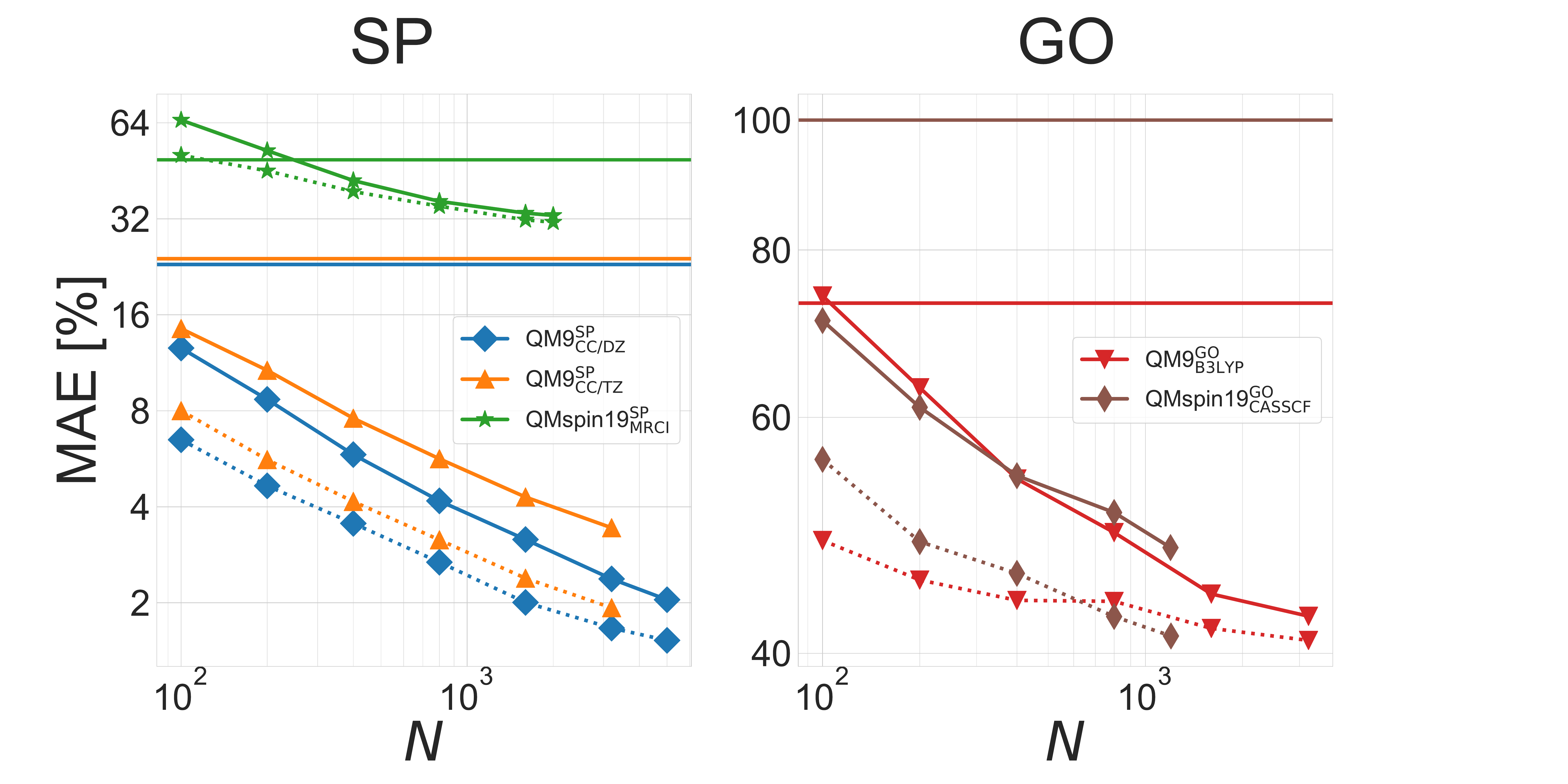}
    \caption{Learning curves showing normalized test errors (cross validated MAE divided by median of test set)  using BoB and FCHL as representations. The model was trained on CPU times. Horizontal lines correspond to the performance assuming all calculations have mean run time (standard deviation divided by the mean wall time of the data set.}
    \label{fig:cpu_cpu}
\end{figure}
\begin{figure}[!ht]
    \centering
    \includegraphics[width=.5\textwidth]{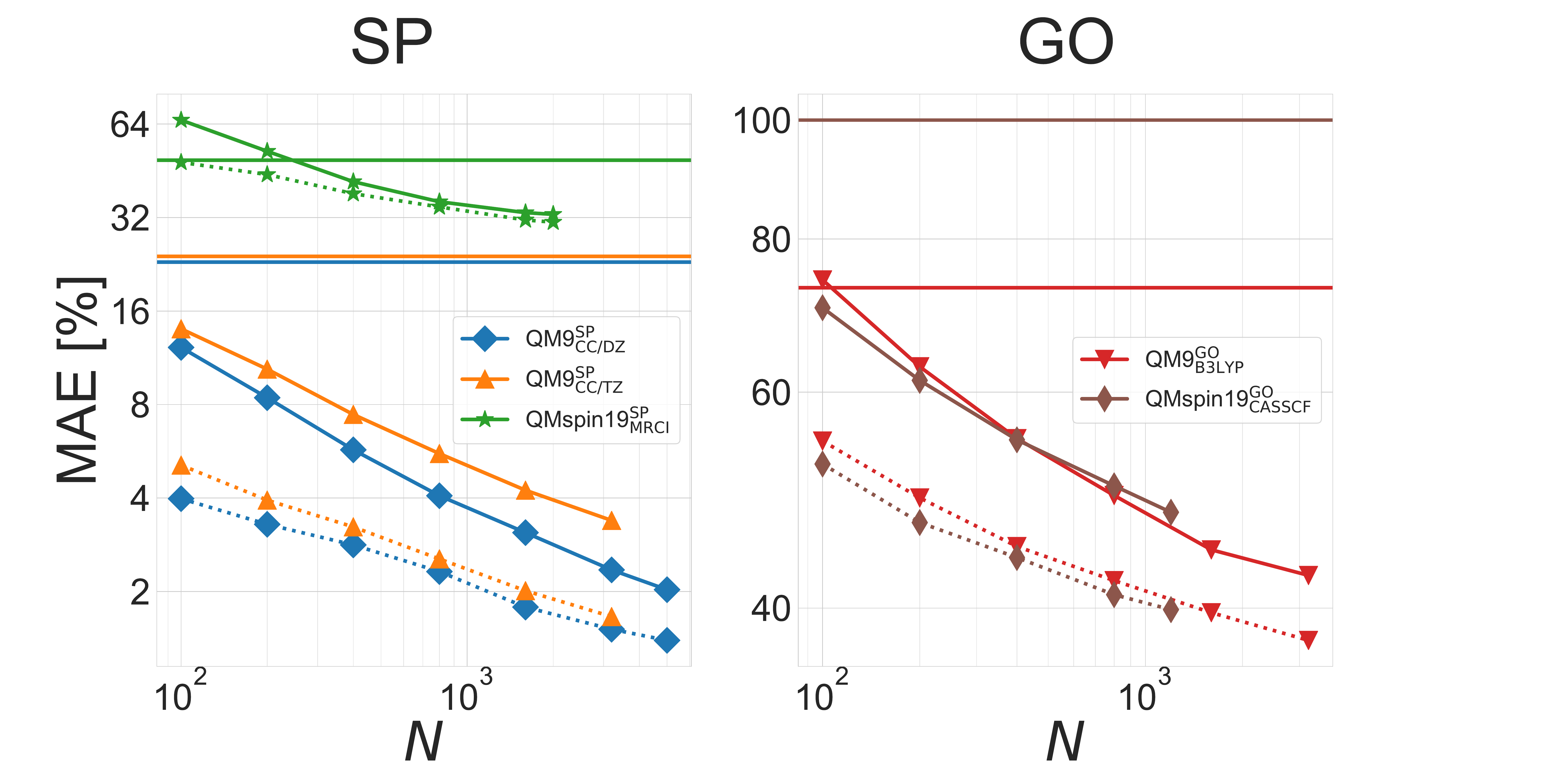}
    \caption{Learning curves showing normalized test errors (cross validated MAE divided by median of test set)  using BoB and FCHL as representations. The model was trained on normalized (by number of electrons) CPU times. Horizontal lines correspond to the performance assuming all calculations have mean run time (standard deviation divided by the mean wall time of the data set.}
    \label{fig:ncpu_cpu}
\end{figure}
\begin{figure}[!ht]
    \centering
    \includegraphics[width=.5\textwidth]{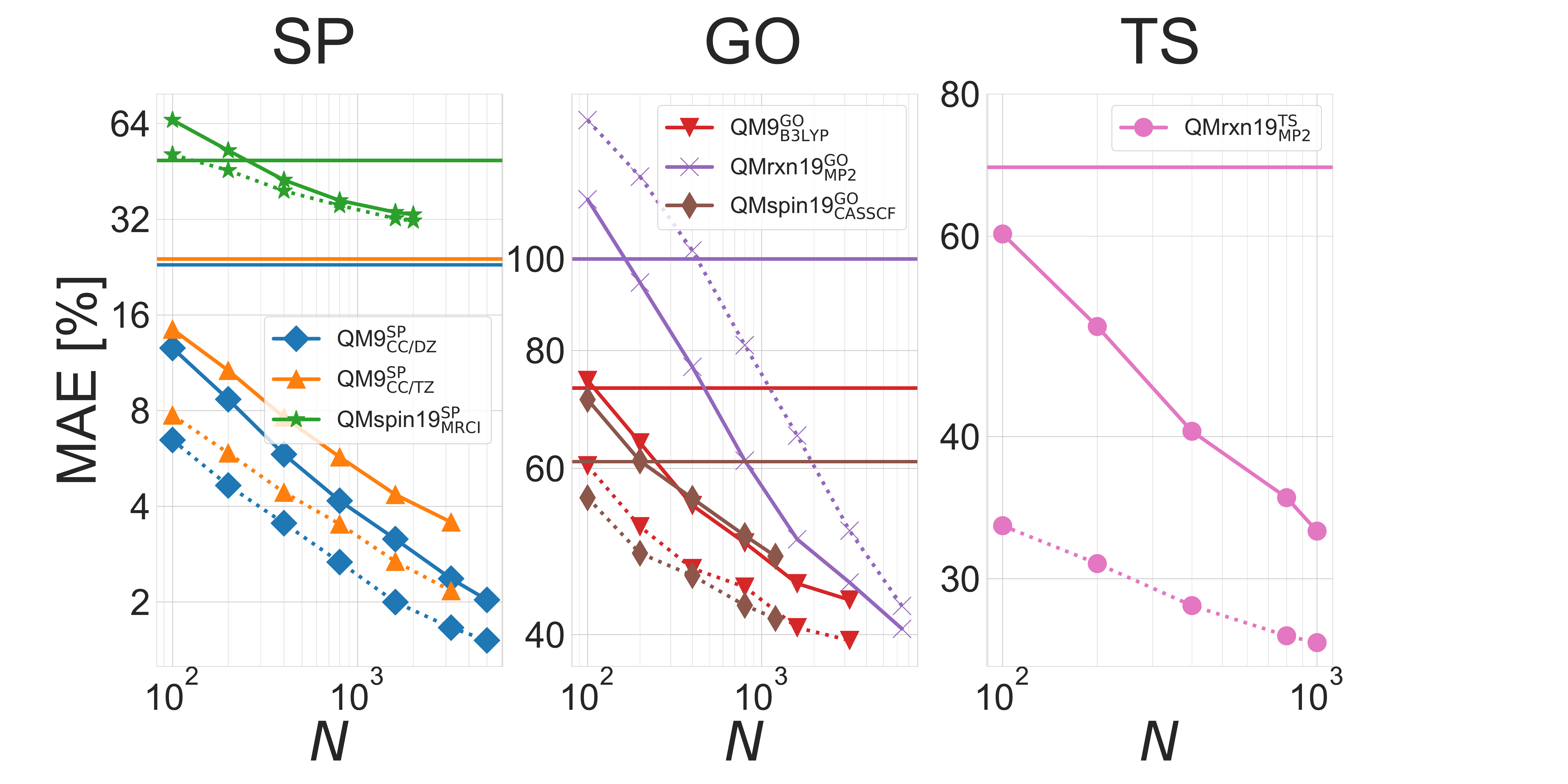}
    \caption{Learning curves showing normalized test errors (cross validated MAE divided by median of test set)  using BoB and FCHL as representations. The model was trained on wall times. Horizontal lines correspond to the performance assuming all calculations have mean run time (standard deviation divided by the mean wall time of the data set.}
    \label{fig:wall_wall}
\end{figure}
\begin{figure}[!ht]
    \centering
    \includegraphics[width=.5\textwidth]{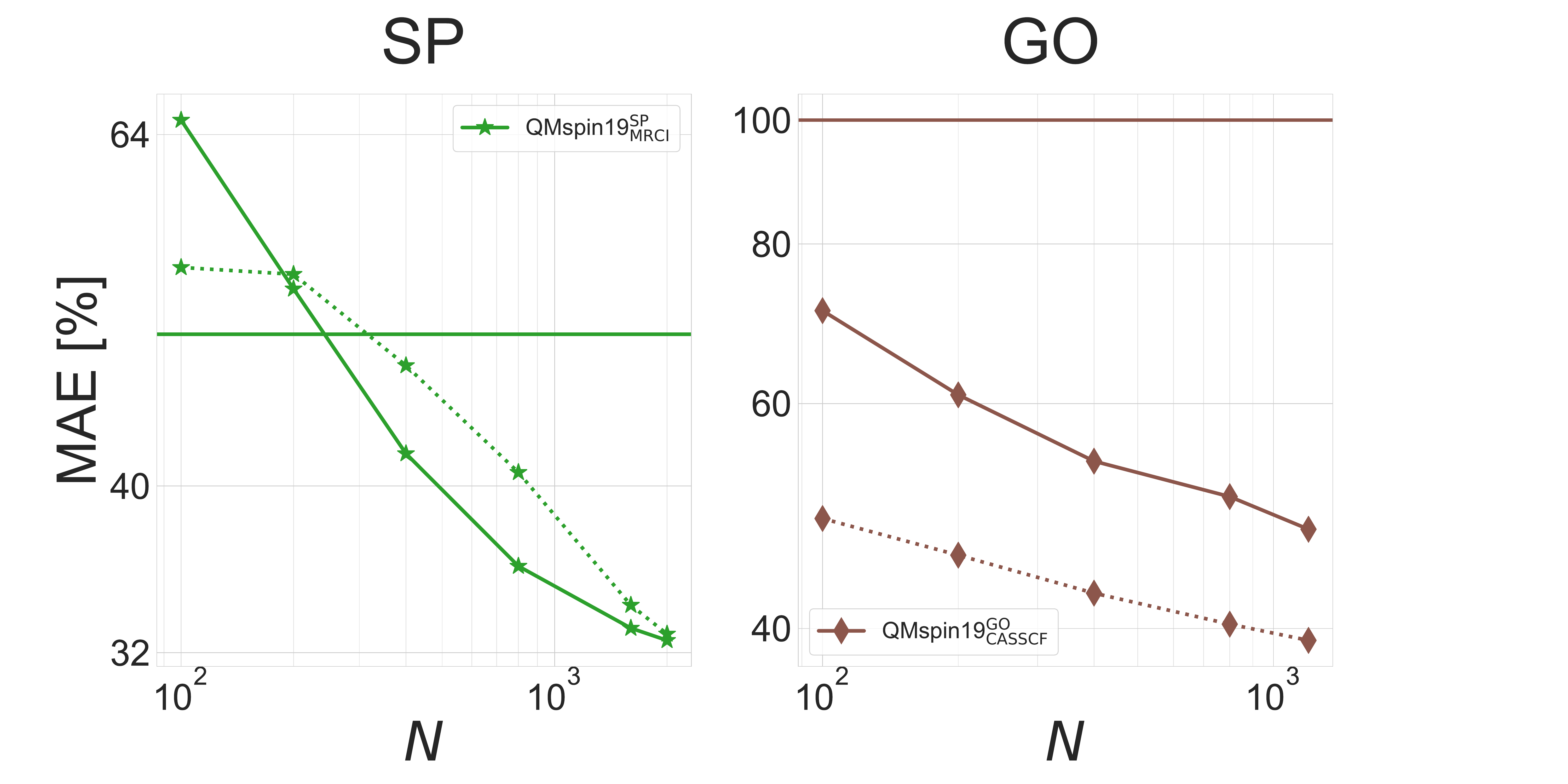}
    \caption{Learning curves showing normalized test errors (cross validated MAE divided by median of test set)  using BoB and FCHL as representations. The model was trained on normalized (number of occupied orbitals to the power of 2 times number of basis functions to the power of 4) wall times. Horizontal lines correspond to the performance assuming all calculations have mean run time (standard deviation divided by the mean wall time of the data set.}
    \label{fig:nmwall_wall}
\end{figure}
\newpage
\bibliography{literatur}